\documentclass[twocolumn,english,pre]{revtex4}
\usepackage[T1]{fontenc}
\usepackage[utf8]{inputenc}
\usepackage{color}
\usepackage{amsmath}
\usepackage{amssymb}
\usepackage{graphicx}
\usepackage{esint}

\makeatletter

\providecommand{\tabularnewline}{\\}

\@ifundefined{textcolor}{}
{%
 \definecolor{BLACK}{gray}{0}
 \definecolor{WHITE}{gray}{1}
 \definecolor{RED}{rgb}{1,0,0}
 \definecolor{GREEN}{rgb}{0,1,0}
 \definecolor{BLUE}{rgb}{0,0,1}
 \definecolor{CYAN}{cmyk}{1,0,0,0}
 \definecolor{MAGENTA}{cmyk}{0,1,0,0}
 \definecolor{YELLOW}{cmyk}{0,0,1,0}
}

\@ifundefined{definecolor}
 {\usepackage{color}}{}
\makeatother

\usepackage{babel}
\begin{document}

\title{{\Large{Occupation Probabilities and Fluctuations in the Asymmetric
Simple Inclusion Process }}}

\author{{\normalsize{Shlomi Reuveni$^{1,2,*}$, Ori Hirschberg$^{3,*}$,
Iddo Eliazar$^{4}$ and Uri Yechiali$^{1}$}}\\
{\normalsize{~}}}

\affiliation{\textit{$^{1}$Department of Statistics and Operations Research,
Tel-Aviv University, Tel-Aviv 69978, Israel}}

\affiliation{\textit{$^{2}$Department of Systems Biology, Harvard University,
Boston, Massachusetts 02115, USA}}

\affiliation{\textit{$^{3}$Department of Physics of Complex Systems, Weizmann
Institute of Science, Rehovot 76100, Israel}}

\affiliation{\textit{$^{4}$Department of Technology Management, Holon Institute
of Technology, Holon 58102, Israel}}

\affiliation{{*} O. Hirschberg and S. Reuveni had equal contribution to this work}
\begin{abstract}
The Asymmetric Simple Inclusion Process (ASIP), a lattice-gas model
of unidirectional transport and aggregation, was recently proposed
as an `inclusion' counterpart of the Asymmetric Simple Exclusion Process
(ASEP). In this paper we present an exact closed-form expression for
the probability that a given number of particles occupies a given
set of consecutive lattice sites. Our results are expressed in terms
of the entries of Catalan's trapezoids --- number arrays which generalize
Catalan's numbers and Catalan's triangle. We further prove that the
ASIP is asymptotically governed by: (i) an inverse square root law
of occupation; (ii) a square root law of fluctuation; and (iii) a
Rayleigh law for the distribution of inter-exit times. The universality
of these results is discussed.
\end{abstract}
\maketitle

\section{\label{sec:Introduction}Introduction}

The Asymmetric Simple Inclusion Process (ASIP) is a unidirectional
lattice-gas flow model which was recently introduced \cite{ASIP-1,Grosskinsky}
as an `inclusion' counterpart of the Asymmetric Simple Exclusion Process
(ASEP) \cite{Derrida1,Derrida2,Golinelli}. In both processes, random
events cause particles to hop uni-directionally, from site to site,
along a one-dimensional lattice. In the ASEP particles are subject
to \emph{exclusion} interactions that keep them singled apart, whereas
in the ASIP particles are subject to \emph{inclusion} interactions
that coalesce them into inseparable particle clusters. The ASIP links
together the ASEP with the Tandem Jackson Network (TJN) \cite{Jackson1,Jackson2,Fundamentals of Queueing Networks}
--- a fundamental service model in queueing theory. From a queueing
perspective, the ASIP's `gluing' of particles into inseparable particle-clusters
manifests unlimited `batch service' \cite{Neuts,Kaspi,van der Wal1,van der Wal2,van der Wal3}
and the model can thus be understood as a TJN with this additional
property. The ASIP is briefly described as follows. Particles enter
a lattice with rate $\lambda$ at its leftmost site, and hop from
one site to the next in clusters. In each hopping event the entire
particle content of a site translocates as one to the next site and
immediately coalesces with the particle content therein. The clusters
continue to hop and coalesce with other clusters until they finally
exit the lattice from its rightmost site. 

Even the simplest ASIPs --- homogeneous ASIPs, in which the hopping
rates do not depend on the position along the lattice --- were shown
to display an intriguing showcase of complexity, including power law
occupations statistics, diverse forms of self-similarity, and a rich
limiting behavior \cite{ASIP-2,ASIP-3}. However, several of the aforementioned
`complexity results' relied only on Monte-Carlo studies, as an exact
expression for the joint stationary probability distribution of particle
occupations is not known. Obtaining an exact, closed form, solution
of the model is undoubtedly difficult, as coalescence introduces strong
correlations between the occupations of different lattice sites. In
Ref. \cite{ASIP-1}, an iterative scheme for the computation of the
probability generating function (PGF) of this steady-state distribution
was presented. And yet, the PGF turns out to be analytically tractable
only for very small ASIPs --- a fact that is manifest in the rapid
growth in its complexity as a function of lattice size \cite{ASIP-1}.
Homogenous ASIPs were nevertheless proven optimal with respect to
various measures of efficiency \cite{ASIP-1} thus further indicating
their special importance. 

The main goal of this paper is to present an exact closed-form expression
for the probability that a given number of particles occupies a given
set of consecutive lattice sites on an homogeneous ASIP lattice. These
probabilities, which we term the incremental load probabilities (to
be defined precisely below), are marginals of the joint occupation
distribution. Progress can be made with their analysis by using the
empty-interval method, a method which has proven useful in the study
of aggregation in closed systems \cite{Kinetic View,Coalescence Process}.
The calculation of these probabilities in our open system is based
on a combinatorial analysis of the incremental load and on the solution
of a boundary value problem that governs its distribution. This approach
yields exact, closed form, results expressed in terms of the entries
of Catalan's trapezoids \cite{Catalan Trapezoid} --- number arrays
which generalize Catalan's numbers and Catalan's triangle \cite{Catalan,Catalan's_triangle1,Catalan's_triangle2,Catalan's_triangle3}. 

The incremental load probabilities provide valuable information on
the ASIP steady state, and furnish an analytical proof for the numerical
results obtained in \cite{ASIP-2}. In particular, we prove that:
(i) the probability that the $k^{th}$ lattice site is non-empty decays
like $1/\sqrt{k}$; (ii) the variance of the occupancy of the $k^{th}$
lattice site grows like $\sqrt{k}$; and (iii) the ASIP's outflow
is governed by Rayleigh-distributed inter-exit times. Thus, in this
paper we present a substantial advance towards the exact solution
of the ASIP model. 

Before presenting the exact expression for the incremental load probabilities,
we follow a complementary approach which is based on mapping the original
problem onto its diffusion limit counterpart. This approach shows
that the incremental load probabilities in lattice segments that are
far away downstream have asymptotic scaling forms which we compute.
Some of these scaling forms were previously found in Ref. \cite{Jain}
using an alternative discrete approach. Here we present a ``real-space''
analysis performed in a continuum limit. Our analysis yields physical
insight into the behavior of the model and allows us to derive some
new asymptotic scaling forms. More importantly, the diffusion-limit
approach reveals that the asymptotics of the incremental load probabilities
are \emph{universal}, in the sense that they do not depend on the
details of the process which feeds particles into the ASIP lattice. 

The paper is organized as follows. Section \ref{sec:Model-Description}
reviews the ASIP, as well as the motivation for introducing this model.
The main results of this paper are summarized in section \ref{sec:Results-Summary}
where we also introduce the notion of incremental load. In section
\ref{sec:The-ASIP-as-a-Coag} the ASIP is described as a coagulation
model and the empty-interval method is adapted to its analysis. In
section \ref{sec:Continuum-limit-of} a continuum diffusion limit
is carried out and asymptotic results are obtained; various implications
of these results are discussed in section \ref{sec:Implications-of-the}.
Section \ref{sec:Incremental-Load-Analysis} further deepens the probabilistic
analysis of the incremental load, and the associated boundary-value
problem. In this section we obtain expressions for the incremental
load which may be efficiently computed even for inhomogeneous ASIPs.
In section \ref{sec:Homogeneous-ASIPs-Exact} we return to homogeneous
systems, for which we solve the boundary value problem and obtain
a set of exact, closed form, results. Section \ref{sec:Conclusion}
concludes the paper with an overview and future outlook. 

A note about notation: Throughout the paper\textbf{ }$\mathbf{\left\langle \xi\right\rangle }$\textbf{
}and $\sigma^{2}\left(\xi\right)$ will denote, respectively, the
mathematical expectation and variance of a real-valued random variable
$\xi$.

\section{\label{sec:Model-Description}The ASIP model}

In this section we briefly review the ASIP. This process was introduced
and explored in \cite{ASIP-1,ASIP-2,ASIP-3} and is described as follows.
Consider a one-dimensional lattice of $n$ sites indexed $k=1,\cdots,n$.
Each site is followed by a gate --- labeled by the site's index ---
which controls the site's outflow. Particles arrive at the first site
($k=1)$ following a Poisson process $\Pi_{0}$ with rate $\lambda$,
the openings of gate $k$ are timed according to a Poisson process
$\Pi_{k}$ with rate $\mu_{k}$ ($k=1,\cdots,n$), and the $n+1$
Poisson processes are mutually independent. Note that from this definition
it follows that the times between particle arrivals are independent
and exponentially distributed with mean $1/\lambda$, and that the
times between the openings of gate $k$ are independent and exponentially
distributed with mean $1/\mu_{k}$ ($k=1,\cdots,n$). A key feature
of the ASIP is its `batch service' property: at an opening of gate
$k$ all particles present at site $k$ transit simultaneously, and
in one batch (one cluster), to site $k+1$ --- thus joining particles
that may already be present at site $k+1$ ($k=1,\cdots,n-1$). At
an opening of the last gate ($k=n$) all particles present at site
$n$ exit the lattice simultaneously. 

Denoting the number of particles present in site $k$ ($k=1,\cdots,n$)
by $X_{k}$, the ASIP's dynamics can be schematically summarized as
follows: (i) first site ($k=1$): 
\begin{equation}
X_{1},X_{2},\cdots\xrightarrow{\lambda}X_{1}+1,X_{2},\cdots\,;\label{201a}
\end{equation}
(ii) interior sites $(1<k\leq n-1)$: 
\begin{equation}
\cdots,X_{k-1},X_{k},X_{k+1},\cdots\xrightarrow{\mu_{k}}\cdots,X_{k-1},0,X_{k+1}+X_{k},\cdots\,;\label{201b}
\end{equation}
(iii) last site ($k=n$):
\begin{equation}
\cdots,X_{n-1},X_{n}\xrightarrow{\mu_{n}}\,\cdots,X_{n-1},0\,.\label{201c}
\end{equation}
Throughout most of the paper we focus on homogeneous ASIPs. In this
subclass of ASIPs, the rates $\{\mu_{k}\}$ --- which are, in general,
different --- are identical: $\mu_{1}=\cdots=\mu_{n}$. 

As was briefly mentioned above, the ASIP is related to several prominent
models both in statistical physics and in queueing theory. We now
review these models and discuss their connection to the ASIP.

\subsubsection{Coagulation-Aggregation }

The ASIP can be viewed as a model of coagulation and aggregation.
Such reaction-diffusion models have been extensively studied since
the pioneering work of Smoluchowski \cite{Smoluchowski}. Yet still,
unresolved issues and intriguing new facets cause them to raise interest
even today \cite{Sokolov,Lindenberg}. Two of the simplest models
of this kind are the coalescence-diffusion model 
\begin{align}
\cdots AA\cdots & \xrightarrow{1}\,\cdots0A\cdots\;,\nonumber \\
\cdots A0\cdots & \xrightarrow{1}\,\cdots0A\cdots\;,\label{203}
\end{align}
where $A$ represents an occupied site and $0$ represents an empty
site, and the aggregation-diffusion model 
\begin{align}
\cdots A_{l}A_{l'}\cdots & \xrightarrow{1}\,\cdots0A_{l+l'}\cdots\;,\nonumber \\
\cdots A_{l}0\cdots & \xrightarrow{1}\,\cdots0A_{l}\cdots\;,\label{204}
\end{align}
where $A_{l}$ represents a site occupied by $l>0$ particles, and
$0$ represents an empty site \cite{Kinetic View}. In both Eqs. (\ref{203})
and (\ref{204}) the rates --- with no loss of generality --- are
set to be one.

The studies dedicated to the models described in Eqs. (\ref{203})
and (\ref{204}) were, by and large, carried out in a one-dimensional
ring topology. Under these conditions many statistical properties
can be calculated exactly using the empty-interval method \cite{Kinetic View,Coalescence Process},
which we shall address in section \ref{sec:The-ASIP-as-a-Coag}. The
ASIP, with homogeneous unit rates $\{\mu_{1}=\cdots=\mu_{n}=1\}$,
can be viewed as a generalization of aggregation-diffusion models
to an open system. Indeed, the bulk ASIP dynamics of Eq. (\ref{201b})
is identical to the dynamics of Eq. (\ref{204}). Similarly, when
one disregards the number of particles occupying each site ($X_{k}$)
and focuses only on whether sites are occupied or not ($X_{k}>0$
or $X_{k}=0$), the ASIP dynamics turns into an open-boundary version
of Eq. (\ref{203}). Previous studies of open-boundary aggregation
models have been carried out in \cite{Derrida3,Jain}. The results
presented herein can be viewed as extensions and generalizations of
these works.

\subsubsection{Asymmetric Simple Exclusion Process}

The ASIP is an exactly solvable `inclusion' counterpart of the Asymmetric
Simple Exclusion Process (ASEP) --- a fundamental model in non-equilibrium
statistical physics \cite{Derrida1,Derrida2,Golinelli}. While both
models share the aforementioned sites-gates lattice structure, the
dynamics of the ASEP is governed by exclusion interactions which do
not allow sites to be occupied by more than a single particle at a
time. To pinpoint the difference between the models consider the two
following characteristic capacities: (i) \emph{site capacity} $c_{site}$
--- the number of particles that can simultaneously occupy a given
site, and (ii) \emph{gate capacity} $c_{gate}$ --- the number of
particles that can be simultaneously transferred through a given gate
when it opens. In the the ASIP $c_{site}=\infty$ and $c_{gate}=\infty$
while in the ASEP $c_{site}=1$ and $c_{gate}=1$. Despite its simple
one-dimensional structure and dynamics, the ASEP displays a complex
and intricate behavior \cite{Derrida1,Derrida2,Golinelli,Blythe}.

The ASEP has a long history having first appeared in the literature
as a model of bio-polymerization \cite{MacDonald} and of stochastic
transport phenomena in general \cite{Spitzer}. Over the years, the
ASEP and its variants were used to study a wide range of physical
phenomena: transport across membranes \cite{Heckmann}, transport
of macromolecules through thin vessels \cite{Levitt}, hopping conductivity
in solid electrolytes \cite{Richards}, reptation of polymer in a
gel \cite{Widom}, traffic flow \cite{Schreckenberg}, gene translation
\cite{Shaw,Reuveni}, surface growth \cite{Halpin,Krug}, sequence
alignment \cite{Bundschuh}, molecular motors \cite{Klumpp} and the
directed motion of tracer particles in the presence of dynamical backgrounds
\cite{Oshanin1,Oshanin2,Oshanin3,Monasterio}.

\subsubsection{Tandem Jackson Network }

Queueing theory is the scientific field focused on the modeling and
analysis of queues \cite{Fundamentals of Queueing Theory}. The `traditional'
applications of queueing theory are common and widespread in telecommunication
\cite{Telecommunications1,Telecommunications2,Telecommunications3},
traffic engineering \cite{Traffic Engineering}, and performance evaluation
\cite{performance1,performance2,performance3}. More recently, some
`non-traditional' applications of queueing theory have attracted interest
--- examples including: human dynamics \cite{human behavior1,human behavior2,human behavior3,human behavior4},
gene expression \cite{gene expression1,gene expression2,Arazi1,Arazi2},
intracellular transport \cite{microtubule}, and non-equilibrium statistical
physics \cite{Kafri,Sandpiles,Arita1,Arita2,Arita3,Chernyak1,Chernyak2}.
The ASIP links together the ASEP with the Tandem Jackson Network (TJN)
\cite{Jackson1,Jackson2,Fundamentals of Queueing Networks} --- a
fundamental service model in queueing theory which represents a sequential
array of Markovian ``single server queues'' \cite{Fundamentals of Queueing Theory}.
In terms of the aforementioned site-capacity and gate-capacity, the
TJN is characterized by: $c_{site}=\infty$ and $c_{gate}=1$. Namely,
each site can accommodate an unlimited number of particles, and at
each gate-opening only one particle can pass through the gate. From
a queueing perspective, the ASIP's `gluing' of particles into inseparable
particle-clusters manifests unlimited `batch service'. Thus, the ASIP
can be viewed as a TJN with the additional property of unlimited batch
service \cite{Neuts,Kaspi,van der Wal1,van der Wal2,van der Wal3}.

\section{\label{sec:Results-Summary}A summary of key results}

In this section we present a short summary of the key results established
in this paper. Some of the results proven herein were previously observed
in numerical simulations \cite{ASIP-2}. In the present work we derive
them analytically and considerably generalize them. In what follows
we consider a homogeneous ASIP with $\mu_{1}=\cdots=\mu_{n}=\mu$,
and set $X_{k}$ to be a random variable which represents the fluctuating
number of particles present in site $k$ in the steady state. We open
this section with a series of asymptotic (large $k$) results for
the distribution and moments of $X_{k}$. The asymptotic results presented
herein all stem from the main result of this paper --- an exact derivation
of the steady-state distribution of the ASIP's incremental load ---
with which we conclude this section.

\emph{Occupation probabilities}. In \cite{ASIP-2} Monte Carlo simulations
concluded that the probability that site $k$ is occupied, $\Pr(X_{k}>0)$,
decays like $1/\sqrt{k}$ (as $k\rightarrow\infty$). Here we analytically
prove that 
\begin{equation}
\Pr(X_{k}>0)=1-\Pr(X_{k}=0)\simeq\frac{1}{\sqrt{\pi k}}\text{ ,}\label{301}
\end{equation}
where ``$\simeq$'' denotes asymptotic equivalence to leading order
in $k$. We further obtain a scaling form for the probability that
site $k$ is occupied by $1\ll l\ll k$ particles: 
\begin{equation}
\Pr(X_{k}=l)\simeq\frac{\mu}{\lambda k}\phi\left(\frac{\mu l}{\lambda\sqrt{k}}\right)\,,\label{302}
\end{equation}
where 
\begin{equation}
\phi(u)=\frac{1}{\sqrt{4\pi}}ue^{-u^{2}/4}.\label{303}
\end{equation}
We note that the result of Eq. (\ref{301}) --- contrary to the result
of Eq. (\ref{302}) --- is universal in the sense that it does not
depend on the arrival rate $\lambda$. In fact, we show that this
result is universal in a wider sense, and that Eq. (\ref{301}) holds
for any particle arrival process (not necessarily Poissonian). A similar,
although slightly weaker, universality holds for the result in Eqs.
(\ref{302})--(\ref{303}): while the scaling variable $u$ depends
on the arrival rate, the scaling function (\ref{303}) does not depend
on the details of the arrival process. The extent to which this claim
is correct is discussed, along with other universality related issues,
in section \ref{sec:Continuum-limit-of}.

\emph{Conditional mean occupancy}. In \cite{ASIP-1} it was shown
that in homogeneous ASIPs the mean occupancy of site $k$ at steady
state is given by 
\begin{equation}
\left\langle X_{k}\right\rangle =\lambda/\mu\label{304}
\end{equation}
 ($k=1,\cdots,n$). Thus, combining the general result of Eq. (\ref{304})
with the result of Eq. (\ref{301}) we obtain that the conditional
mean occupancy of site $k$, conditioned on the information that the
site is not empty, is given by 
\begin{equation}
\langle X_{k}|X_{k}>0\rangle\simeq\frac{\lambda}{\mu}\sqrt{\pi k}\text{ .}\label{305}
\end{equation}
The power-law asymptotics of Eqs. (\ref{301}) and (\ref{305}) imply
that the stationary occupation of `downstream' sites (large $k$)
exhibits large fluctuations. On the one hand, a downstream site is
rarely occupied: $\Pr(X_{k}>0)\simeq1/\sqrt{\pi k}$. On the other
hand, when a downstream site is occupied then its conditional mean
is dramatically larger than its mean --- the former being of order
$O(\sqrt{k})$, while the latter being of order $O\left(1\right)$. 

\emph{Fluctuations}. A square root law of fluctuation, in which the
variance in the occupancy of site $k$ grows like $\sqrt{k}$, was
numerically observed in \cite{ASIP-2}. Here we prove that 
\begin{equation}
\sigma^{2}(X_{k})\simeq\frac{4\lambda^{2}}{\mu^{2}}\sqrt{\frac{k}{\pi}}\text{ .}\label{306}
\end{equation}
Equation (\ref{306}) is obtained by substituting Eq. (\ref{302})
into the second moment $\langle X_{k}^{2}\rangle=\overset{\infty}{\underset{l=1}{\sum}}l^{2}\Pr(X_{k}=l)$,
approximating the second moment by a corresponding integral, and noting
that $\sigma^{2}(X_{k})\simeq\langle X_{k}^{2}\rangle$ (as the mean
$\langle X_{k}\rangle$ is constant in $k$). 

\emph{Inter-exit times}. Consider the times at which particle clusters
exit site $k$, and let $T_{k}$ denote the time elapsing between
two such consecutive exit events at steady state. Here we prove that
the probability density of the scaled inter-exit time $T_{k}/\sqrt{\pi k}$
is asymptotically governed by the Rayleigh distribution 
\begin{equation}
P_{T_{k}/\sqrt{\pi k}}(t)\simeq\frac{\pi t}{2}\exp\left(-\pi t^{2}/4\right)\label{308}
\end{equation}
$(t>0)$, as previously anticipated by Monte-Carlo simulations \cite{ASIP-2}. 

\emph{Incremental load.} The ASIP's overall load is the total number
of particles present in the lattice at steady state. The steady state
distribution of the overall load was comprehensively analyzed in \cite{ASIP-1}.
Generalizing the concept of the overall load we consider a `lattice
interval', contained within the ASIP lattice, which starts at site
$k$ and consists out of $m$ consecutive sites: $\{k,k+1,\cdots,k+m-1\}$
($k,m=1,2,3,\cdots$). The ASIP's incremental load corresponding to
this lattice interval at steady state is given by
\begin{equation}
L\left(k,m\right)=\sum_{i=k}^{k+m-1}X_{i}\text{ .}\label{309}
\end{equation}
Clearly, the number of particles occupying site $k$, $L(k,1)$, and
the overall load, $L(1,n)$, are both special cases of the incremental
load $L(k,m)$. The main result of this paper is an exact closed-form
expression for the distribution of the incremental load
\begin{equation}
P_{l}(k,m)\equiv\text{Pr}\big(L(k,m)=l\big)\,,\label{310}
\end{equation}
 ($l=0,1,2,\cdots$). This expression, presented in Eq. (\ref{808a}),
is given in terms of the entries of Catalan's trapezoids \cite{Catalan Trapezoid}.

\section{\label{sec:The-ASIP-as-a-Coag}The ASIP as a coagulation model }

As discussed in Section \ref{sec:Model-Description}, coagulation
models similar to the ASIP have been analyzed successfully using the
empty-interval method and its generalization to non-empty intervals.
In this method, one studies the steady state distribution of the incremental
load defined in Eq. (\ref{309}), and the time evolution of its associated
time-dependent counterpart 
\begin{equation}
L\left(t;k,m\right)=\sum_{i=k}^{k+m-1}X_{i}(t)\text{ ,}\label{401}
\end{equation}
where $X_{i}(t)$ denotes the number of particles present in site
$i$ at time $t$ ($t\geq0$). In this section we review the method
and show how it is applied to the analysis of the ASIP. 

We begin with the probability that the lattice interval $\{k,k+1,\cdots,k+m-1\}$
is empty at time $t$:
\begin{equation}
P_{0}(t;k,m)\equiv\text{Pr}\bigl(L(t;k,m)=0\bigr).\label{402}
\end{equation}
The empty-interval method is based on the fact that it is possible
to write a closed-form evolution equation for the probabilities $P_{0}(t;k,m)$
as follows. 

Consider a homogeneous ASIP. By rescaling time, the homogeneous gate
opening rate and the particle arrival rate can be normalized to $\mu\rightarrow1$
and $\lambda\rightarrow\lambda/\mu$ correspondingly. Accordingly,
from this point onward we will assume, without loss of generality,
that $\mu=1$ and that $\lambda$ is measured in units of the gate
opening rate. For $k>1$ and $m>1$, the probability $P_{0}(t;k,m)$
evolves according to the equation 
\begin{equation}
\begin{array}{l}
\frac{\partial}{\partial t}P_{0}(t;k,m)=\left[P_{0}(t;k,m-1)-P_{0}(t;k,m)\right]\\
\\
-\bigl[P_{0}(t;k,m)-P_{0}(t;k-1,m+1)\bigr]\,.\\
\text{ }
\end{array}\label{403}
\end{equation}
The term $P_{0}(t;k,m-1)-P_{0}(t;k,m)$ appearing on the right-hand
side of Eq. (\ref{403}) manifests the probability that sites $\{k,k+1,\cdots,k+m-2\}$
are empty and site $k+m-1$ is occupied, in which case the particle
cluster at site $k+m-1$ might hop (with rate 1) to site $k+m$ and
thus leave the interval $\{k,\cdots,k+m-1\}$ empty, as illustrated
in Figure 1. Similarly, the term $P_{0}(t;k,m)-P_{0}(t;k-1,m+1)$
appearing on the right-hand side of Eq. (\ref{403}) manifests the
probability that sites $\{k,k+1,\cdots,k+m-1\}$ are empty and site
$k-1$ is occupied, in which case the particle cluster at site $k-1$
might hop to site $k$ (with rate 1), thus rendering the interval
$\{k,k+1,\cdots,k+m-1\}$ non-empty as illustrated in Figure 2. 
\begin{figure}
\begin{centering}
\includegraphics[scale=0.45]{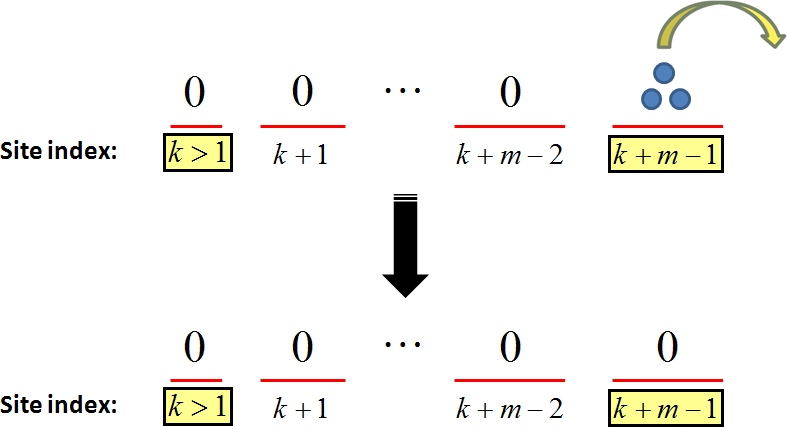}
\par\end{centering}

\caption{(color online). The non-empty interval $\{k,\ldots,k+m-1\}$ becomes
empty if, and only if, all interval sites other than site $k+m-1$
are empty and the particles that occupy site $k+m-1$ hop to site
$k+m$. }
\end{figure}
\begin{figure}
\begin{centering}
\includegraphics[scale=0.45]{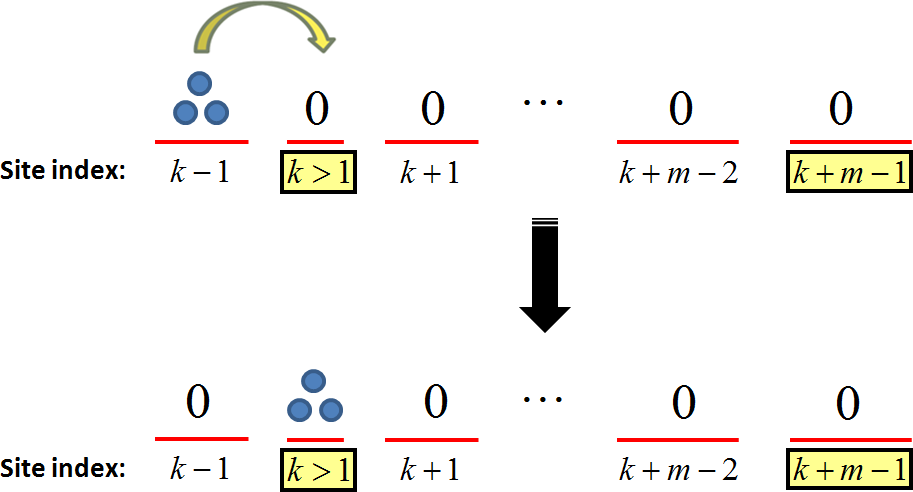}
\par\end{centering}

\caption{(color online). The empty interval $\{k,\ldots,k+m-1\}$ becomes non-empty
if, and only if, site $k-1$ is occupied and the particles that occupy
it hop to site $k$. }
\end{figure}

Eq. (\ref{403}) remains valid for $m=1$ and $k>1$ provided that
we impose the boundary condition 
\begin{equation}
P_{0}(t;k,0)\equiv1\,,\label{405}
\end{equation}
i.e., degenerate intervals (which contain no sites) are by convention
always empty. For $k=1$ and $m\geq1$ the evolution is given by {\small{
\begin{equation}
\frac{\partial}{\partial t}P_{0}(t;1,m)=\left[P_{0}(t;1,m-1)-P_{0}(t;1,m)\right]-\lambda P_{0}(t;1,m)\,.\label{406}
\end{equation}
}}The term $P_{0}(t;1,m-1)-P_{0}(t;1,m)$ appearing on the right-hand
side of Eq. (\ref{406}) manifests the probability that sites $\{1,2,\cdots,m-1\}$
are empty and site $m$ is occupied, in which case the particle cluster
at site $m$ might hop (with rate 1) to site $m+1$ and thus leave
the interval $\{1,\cdots,m\}$ empty. Also, $P_{0}(t;1,m)$ is the
probability that the interval $\{1,\cdots,m\}$ is empty, in which
case a particle might arrive to site $1$ (with rate $\lambda$),
thus rendering the interval $\{1,\cdots,m\}$ non-empty.

The empty-interval method can be generalized to capture the evolution
of the probability $P_{l}(t;k,m)$ that there are exactly $l$ particles
at sites $\{k,k+1,\cdots,k+m-1\}$ at time $t$ \cite{Coalescence Process,Kinetic View}:
\begin{equation}
P_{l}(t;k,m)\equiv\text{Pr}\bigl(L(t;k,m)=l).\label{408}
\end{equation}
The empty-interval probabilities $P_{0}(t;k,m)$ are hence a special
case of $P_{l}(t;k,m)$ with $l=0$. The counterparts of Eqs. (\ref{403})--(\ref{406})
are as follows (see Appendices \ref{sub:Derivation-of-Eq. 409} and
\ref{sub:Derivation-of-Eq. 411} for the derivations). For $k>1$
and $m>1$ the evolution is given by 
\begin{equation}
\begin{array}{l}
\frac{\partial}{\partial t}P_{l}(t;k,m)=\\
\\
+\Bigl[P_{l}(t;k,m-1)-2P_{l}(t;k,m)+P_{l}(t;k,m+1)\Bigr]\\
\\
-\Bigl[P_{l}(t;k,m+1)-P_{l}(t;k-1,m+1)\Bigr]\,.\\
\text{ }
\end{array}\label{409}
\end{equation}
Equation (\ref{409}) remains valid for $m=1$ and $k>1$ provided
that we impose the boundary condition 
\begin{equation}
P_{l}(t;k,0)=\delta_{l,0}\,.\label{410}
\end{equation}
where $\delta_{l,0}$ is the Kronecker delta symbol. Note that, remarkably,
Eqs. (\ref{409}) for $P_{l}(t;k,m)$ do not couple different values
of $l$. A coupling enters only through the boundary condition $P_{l}(t;1,m)$
($m\geq1$) whose time evolution is given by
\begin{equation}
\begin{array}{l}
\frac{\partial}{\partial t}P_{l}(t;1,m)=\\
\\
+\Bigl[P_{l}(t;1,m-1)-P_{l}(t;1,m)\Bigr]\\
\\
-\lambda\Bigl[P_{l}(t;1,m)-P_{l-1}(t;1,m)\Bigr]\,.\\
\text{ }
\end{array}\label{411}
\end{equation}
Note that setting $l=0$ in Eqs. (\ref{409})--(\ref{411}), while
taking into account that the probability to observe a negative number
of particles is zero by definition, indeed yields Eqs. (\ref{403})--(\ref{406}).

\section{\label{sec:Continuum-limit-of}Continuum limits of the steady-state
equations}

The main result of this paper is an exact expression for the steady-state
solution of Eqs. (\ref{403})--(\ref{411}). Before presenting and
deriving this exact solution (see Sections \ref{sec:Incremental-Load-Analysis}
and \ref{sec:Homogeneous-ASIPs-Exact}) we provide in the current
section a derivation of the asymptotic scaling forms that this solution
attains for large values of $k$, i.e., for lattice intervals located
far away downstream. As discussed above, some of the asymptotic results
presented in this section have been obtained before in \cite{Jain}
using Laplace transform methods. Here we present an alternative ``real-space''
derivation, which yields new physical insight into the solutions and
highlights their universal nature. 

The asymptotic analysis of Eqs. (\ref{403})--(\ref{411}) is based
on the following continuum-limit assumption: if the steady state probability
$P_{l}(k,m)$ changes slowly as a function of the variables $k$ and
$m$, then this discrete function may be approximated by one which
is continuous both in $k$ and $m$. Thus, one can expand to leading
order all terms in the equation around $P_{l}(k,m)$. In this continuum
limit, the discrete Laplacian in the first square brackets of Eq.
(\ref{409}) approximately equals a continuous Laplacian, and similarly
the second square brackets is approximately $\frac{\partial}{\partial k}P_{l}(k,m)$.
Therefore, in the steady-state, where the left-hand side of Eq. (\ref{409})
vanishes, one finds that $P_{l}(k,m)$ satisfies a diffusion equation
where the site number $k$ plays the role of time: 
\begin{equation}
\frac{\partial}{\partial k}P_{l}(k,m)=\frac{\partial^{2}}{\partial m^{2}}P_{l}(k,m)\,.\label{501a}
\end{equation}
This continuum approximation will be shown \textit{a-posteriori} to
be valid when $k\gg m$. 

Equation (\ref{501a}) should be solved with the appropriate boundary
conditions in ``space'' (i.e., in $m$) and ``time'' (i.e., in
$k$). The spatial ($m=0$) boundary condition of Eq. (\ref{501a})
is given in Eq. (\ref{410}), $P_{l}(k,0)=\delta_{l,0}$. The temporal
($k=1$) initial condition is the steady state solution of Eq. (\ref{411}),
which was found to be \cite{ASIP-1}: 
\begin{equation}
P_{l}(1,m)=\binom{l+m-1}{l}\biggl(\frac{1}{1+\lambda}\biggr)^{m}\biggl(\frac{\lambda}{1+\lambda}\biggr)^{l}\,.\label{501c}
\end{equation}

Before proceeding with the study of Eq. (\ref{501a}), let us discuss
its relation with the behavior of an ASIP on a ring. Unlike the open
boundary ASIP on which we focus, on a ring the steady-state behavior
of the model is trivial: a single occupied site circulates throughout
the system uni-directionally. The relaxation to this steady-state,
however, has an interesting scaling form which has been studied extensively
in the context of coagulation models (\ref{203})--(\ref{204}). In
particular, it is known that in a spatially homogeneous ring, the
probability to find $l$ particles in an interval of $m$ sites evolves
(in a continuum limit) according to the diffusion equation (\ref{501a})
with $k$ replaced by time. In other words, as one progresses from
left to right along a stationary open-boundary ASIP, the probability
to see empty or occupied intervals changes (in space) just like the
temporal evolution of the corresponding probability on a ring. This
mapping between the two problems provides an interesting physical
picture: it suggests that the open-boundary ASIP can be thought of
as a sort of a ``conveyor belt'', along which the coagulation reaction
proceeds. A single steady-state snapshot of the open-boundary ASIP
is, in this sense, similar to the entire temporal evolution of the
coagulation model on a ring.

It is well known that the diffusion equation on an infinite line has,
at times which are large compared with (the square of) the spatial
extent of the initial condition, a scaling form of a spreading Gaussian.
Having arrived at the diffusion equation (\ref{501a}), it is not
too surprising that a similar scaling solution is found for it at
large $k$. This solution, however, is not Gaussian, due to the boundary
condition (\ref{410}) which is either a source at the origin when
$l=0$ or a sink when $l\geq1$. In Subsections \ref{sub:ScalingL0}
and \ref{sub:ScalingLnot0} below, we separately describe and derive
the scaling solutions for these two cases. A third, somewhat more
subtle, scaling solution is found when considering the joint limit
of $l\sim\sqrt{k}\gg1$. In this case, $k$ is not large enough in
comparison with the spatial extent of the ``initial condition''
(\ref{501c}) in order for the usual scaling of the diffusion equation
to apply. Nonetheless, $P_{l}(k,m)$ is found to have a universal
scaling form in the variable $l/\sqrt{k}$. This scaling form is discussed
in Subsection \ref{sub:ScalingLlikeSqrtK}. The universality of the
obtained scaling forms and the conditions under which the continuum
approximation is valid are discussed in Subsection \ref{sub:ScalingRemarks}.

\subsection{The case of $l=0$\label{sub:ScalingL0}}

As with the usual (probability conserving) diffusion in its late stages,
the large $k$ solution of Eq. (\ref{501a}) is given by a scaling
form. This form can be found by substituting the ansatz 
\begin{equation}
P_{l}(k,m)=k^{-\beta}f\left(\frac{m}{\sqrt{k}}\right)\label{502a}
\end{equation}
in Eq. (\ref{501a}), yielding the ordinary differential equation
\begin{equation}
f''(u)+\frac{u}{2}f'(u)+\beta f(u)=0\label{502b}
\end{equation}
for the scaling function $f(u)$, where $u=m/\sqrt{k}$ is the corresponding
scaling variable. 

In the case of $l=0$ (i.e., the probability to see empty intervals),
the boundary condition $P_{0}(k,0)=1$ implies that $\beta=0$ and
$f(0)=1$. The solution of Eq. (\ref{502b}) with this boundary condition
is given by $f(u)=1+C\,\textrm{erf}(u/2)$ where $C$ is an integration
constant, and $\textrm{erf}(x)\equiv2/\sqrt{\pi}\int_{0}^{x}\exp(-y^{2})dy$
is the error function. For large $u$ this solution approaches $1+C$.
Since $\underset{m\to\infty}{\lim}P_{0}(k,m)\to0$ (i.e., there is
a vanishing probability that all sites from $k$ onwards are empty),
the constant $C$ must equal $-1$, yielding the scaling solution
$f(u)=\textrm{erfc}(u/2)$, i.e., 
\begin{equation}
P_{0}(k,m)\simeq\textrm{erfc}\Bigl(\frac{m}{2\sqrt{k}}\Bigr),\qquad(m\ll k)\label{502c}
\end{equation}
where $\textrm{erfc}$ is the complementary error function defined
as $\textrm{erfc}(x)\equiv1-\textrm{erf}(x)$. Here and in the next
two Subsections we indicate in brackets the limiting regime in which
the obtained scaling solutions are valid. These are explained below
in Subsection \ref{sub:ScalingRemarks}.

\subsection{The case of $1\leq l\ll\sqrt{k}$\label{sub:ScalingLnot0}}

When $l\geq1$, Eq. (\ref{501a}) should be solved under the absorbing
boundary condition $P_{l}(k,0)=0$, which by use of Eq. (\ref{502a})
implies that $f(0)=0$. The corresponding solution of Eq. (\ref{502b})
is 
\begin{equation}
f(u)=C\, u\,_{1}F_{1}(\beta+1/2;3/2;-u^{2}/4),\label{503a}
\end{equation}
where $C$ is once again an integration constant and $_{1}F_{1}(a;b;z)$
is the Kummer hypergeometric function. The values of $\beta$ and
$C$ can be determined by using the fact that the quantity $\Lambda=\int_{0}^{\infty}mP_{l}(k,m)dm$
is conserved by the diffusion equation (\ref{501a}) with an absorbing
boundary condition, i.e., it can be shown that $d\Lambda/dk=0$ \cite{Barenblatt}.
The discrete counterpart of this conservation law, which results from
Eq. (\ref{409}), states that 
\begin{equation}
\Lambda_{l}\equiv\sum_{m=1}^{\infty}(m-1)P_{l}(k,m)\label{503b}
\end{equation}
is independent of $k$ in the steady state. For the scaling solution
given by the combination of Eqs. (\ref{502a}) and (\ref{503a}),
$\Lambda_{l}\simeq k^{1-\beta}\int uf(u)du=k^{1-\beta}\sqrt{4\pi}C,$
and we therefore find that $\beta=1$, for which $f(u)=Cu\exp(-u^{2}/4)$
{\small{\cite{Abramowitz}}}, and $C=\Lambda_{l}/\sqrt{4\pi}$, i.e.,
{\small{
\begin{equation}
P_{l}(k,m)\simeq\frac{\Lambda_{l}m}{\sqrt{4\pi}k^{3/2}}e^{-\frac{m^{2}}{4k}}\quad\;(1\leq l\ll\sqrt{k};\, m\ll k).\label{503c}
\end{equation}
}}The value of $\Lambda_{l}$ is found from the initial condition
(\ref{501c}) to be 
\begin{equation}
\Lambda_{l}=\sum_{m=1}^{\infty}(m-1)P_{l}(1,m)=(l+1)/\lambda^{2}.\label{503d}
\end{equation}
To see this, note that up to a multiplication by $\lambda^{-1}$,
Eq. (\ref{501c}) is the probability mass function of a sum of $l+1$
independent geometric random variables with mean $\lambda^{-1}$. 

Note that the scaling form (\ref{503c}) is valid only in the asymptotic
regime when the diffusive length $\sqrt{k}$ is much larger than the
spatial spread of the initial condition, which in our case is of the
same order of $\Lambda_{l}$. In other words, for any fixed $l\geq1$,
Eq. (\ref{503c}) is a good approximation at ``times'' where $\sqrt{k}\gg l$.
In the next Subsection we examine what happens at ``times'' $\sqrt{k}\sim l$
which are not large enough for the initial condition to be washed
out by the diffusion.

\subsection{The case of $l\sim\sqrt{k}$\label{sub:ScalingLlikeSqrtK}}

When $l\sim\sqrt{k}$ and $k$ is not large enough for the diffusion
to reach its asymptotic scaling regime, there seems to be no a-priori
reason to expect a scaling solution to Eq. (\ref{501a}). However,
a closer inspection of the initial condition (\ref{501c}) reveals
that such a scaling solution does exist and, surprisingly, is also
universal. We now derive this scaling solution; its universality is
discussed in the next Subsection. 

The key observation now is that the dependence on the number of particles
$l$ enters only through the initial condition of Eq. (\ref{501c}),
which in the limit we study, and as a function of $m$, is narrowly
centered around $m\simeq l/\lambda$. This once again follows from
the fact that the initial condition of Eq. (\ref{501c}) is proportional
to the probability mass function of a sum of $l+1$ independent geometric
random variables with mean $\lambda^{-1}$. Therefore, according to
the central limit theorem, the distribution of this sum can be approximated,
when $l\to\infty$, by a Gaussian distribution whose mean is given
by $\langle m\rangle=(l+1)/\lambda\simeq l/\lambda$. Recalling that
the standard deviation scales as $\sqrt{l}$, and is therefore negligible
with respect to the mean, we can further approximate the Gaussian
probability density function by a Dirac $\delta$ function, i.e.,
\begin{equation}
P_{l}(1,m)\simeq\lambda^{-1}\,\delta(m-l/\lambda).\label{505}
\end{equation}

The solution of the diffusion equation (\ref{501a}) with an absorbing
boundary at the origin and the initial condition (\ref{505}) is found
(e.g., by the method of images \cite{Images}) to be 
\begin{multline}
\begin{array}{l}
P_{l}(k,m)\simeq\frac{1}{\sqrt{4\pi\lambda^{2}k}}\Bigl[e^{-\frac{(m-l/\lambda)^{2}}{4k}}-e^{-\frac{(m+l/\lambda)^{2}}{4k}}\Bigr]\\
\text{ }
\end{array}\\
(1\ll l\ll k;\, m\ll k).\label{506}
\end{multline}
Equation (\ref{506}) is a joint scaling solution in the scaling variables
$m/\sqrt{k}$ and $l/\sqrt{k}$. If one is further interested in the
limit of $m\ll l$, one may expand and obtain to leading order a ``thermal
dipole'' 
\begin{equation}
P_{l}(k,m)\simeq\frac{ml}{\sqrt{4\pi}\lambda^{2}k^{3/2}}e^{-\frac{l^{2}}{4\lambda^{2}k}}\qquad(m\ll l\ll k).\label{507}
\end{equation}
Note that, as explained below, Eqs. (\ref{506}) and (\ref{507})
are valid not only at the scale of $l\sim\sqrt{k}$, but in fact for
all $1\ll l\ll k$.

\subsection{Remarks on the scaling solutions\label{sub:ScalingRemarks}}

In this Subsection we remark on the limits of validity of the scaling
solutions obtained above, and discuss their universality.

The validity of the scaling solutions obtained in the previous sections
relies on the continuum approximation of the exact (discrete) Eq.
(\ref{409}) by the continuous Eq. (\ref{501a}). A straightforward
calculation shows that the solutions (\ref{502c}), (\ref{503c}),
(\ref{506}), and (\ref{507}) satisfy 
\begin{equation}
P_{l}(k,m+1)-P_{l}(k-1,m+1)=\frac{\partial}{\partial k}P_{l}(k,m)\biggl[1+O\Bigl(\frac{m}{k},\frac{l}{k}\Bigr)\biggr].
\end{equation}
and similarly for the discrete $m$-Laplacian. Therefore, the continuum
approximation is valid as long as ${m,l\ll k}$. Note in particular
that the continuum limit does not require $m$ to be large, and thus
the results are valid even for $m=1$.

An important feature of the scaling solutions (\ref{502c}), (\ref{503c}),
(\ref{506}), and (\ref{507}) is their \emph{universality} with respect
to the details of the how particles arrive at the first site: while
the arrival process dictates the distribution of $L(1,m)$, i.e.,
the initial condition $P_{l}(1,m)$, the scaling solutions are rather
insensitive to it. In other words, one may say that the arrival process
which feeds particles into the ASIP ``conveyor belt'' does not affect
the load statistics far away downstream. As discussed shortly, universality
breaks down for some exotic initial conditions with fat tails, but
is otherwise expected to hold for a rather large class of arrival
processes.

For the scaling solutions (\ref{502c}) and (\ref{503c}), the origin
of universality is easily understood from the diffusion picture of
Eq. (\ref{501a}): it is well known that solutions of the diffusion
equation converge at late times to scaling functions that are independent
of the initial condition (as long as the tail of the initial condition
decay rapidly enough) \cite{Barenblatt}. We now note that $P_{0}(1,m)\leq P_{0}(2,m-1)\leq1/2^{m-1}$
for any arrival process, as can be clearly seen by considering a limiting
scenario in which the arrival process is such that the first site
is always occupied. Hence, the initial condition $P_{0}(1,m)$ decays
(at least) exponentially fast in $m$ and the pathological case of
heavy tails is excluded. As a result, Eq. (\ref{502c}) is not only
independent of $\lambda$ in the case of Poissonian arrivals but also
completely insensitive to nature of the arrival process altogether. 

Equation (\ref{503c}) is also universal, except for the prefactor
$\Lambda_{l}$ given by Eq. (\ref{503d}). This prefactor (and only
it) depends on the details of the arrival process, and is thus non
universal. However, when $l\gg1$ and for initial conditions which
can be approximated by Eq. (\ref{505}) (see discussion shortly),
the prefactor attains the universal form $\Lambda_{l}\simeq l/\lambda^{2}$.
This form still ``remembers'' the mean arrival rate $\lambda$,
but is otherwise independent of the arrival process. Its dependence
on $\lambda$ is both mathematically unavoidable, due to the conservation
of $\Lambda_{l}$, and physically reasonable, as the mean number of
particles per site in Eq. (\ref{304}) depends on $\lambda$. The
universality of Eq. (\ref{503c}) breaks down for fat-tailed $P_{l}(1,m)$
for which $\Lambda_{l}$ diverges. 

The universality of Eqs. (\ref{506}) and (\ref{507}) has a somewhat
more subtle origin. As explained above, these scaling forms are valid
even though $k$ is \emph{not} large enough to ``wash out'' the
initial condition $P_{l}(1,m)$. Rather, they emerge exactly when
the diffusive length $\sqrt{k}$ is of the order of the initial length
scale $\sum_{m}mP_{l}(1,m)\sim l$.\textcolor{red}{{} }The validity
of these scaling functions rests on the approximation in Eq. (\ref{505}),
which itself is a result of the central limit theorem. Therefore,
the scaling forms (\ref{506}) and (\ref{507}) hold whenever the
arrival process is such that $L(1,m)$ lends itself to one of the
many extensions and generalizations of the central limit theorem.
This universality is demonstrated by a specific exactly-solvable example
in Appendix \ref{sec:AppendixUniversality}. 

The scaling forms (\ref{506}) and (\ref{507}) will hold even when
the central limit theorem breaks down, and as long as the standard
deviation in $L(1,m)$ is negligible with respect to its mean in the
limit of $m\rightarrow\infty$. When this is the case, the distribution
of $L(1,m)$ is sharply peaked around its mean thus asserting the
existence of an approximation of the type appearing in Eq. (\ref{505}).
The basin of attraction for this type of behavior is very large. Indeed,
for a general arrival process, Little's law \cite{Wolff} asserts
that $\left\langle L(1,m)\right\rangle =\bar{\lambda}m$, where $\bar{\lambda}$
is the effective arrival rate (long term average of the number of
particles arriving per unit time) and $m$ is the average time a particle
spends in the system. On the other hand, fluctuations in $L(1,m)$
are only caused by arrivals to the first site and departures from
the last site. And so, given the universality of Eq. (\ref{502c}),
if the typical fluctuation due to an arrival event is finite and when
$m$ is large, fluctuations in $L(1,m)$ will be dominated by departure
events. Hence, the standard deviation in $L(1,m)$ will be of order
$\sqrt{m}$ and, most importantly, negligible with respect to the
mean.

\section{\label{sec:Implications-of-the}Implications of the inter-particle
distribution function}

In this section we use the results of Section \ref{sec:Continuum-limit-of}
to derive the scaling properties of the ASIP which were presented
in Section \ref{sec:Results-Summary}.

\emph{Occupation probabilities}. We begin by examining the probability
that a site is occupied. Substituting $m=1$ in Eq. (\ref{502c})
and expanding to first order in $k$, we recover Eq. (\ref{301}).
The occupation-number distribution of a single site, $P_{l}(k,1)$,
is found by substituting $m=1$ in Eq. (\ref{507}). Recalling that
we have rescaled time such the $\mu=1$, we recover the scaling form
reported in Eqs. (\ref{302}) and (\ref{303}). In fact, combining
(\ref{503c}) with (\ref{507}) we may write a uniform approximation
which is asymptotically exact for all $l\geq1$ in the limit of $k\gg1$:

\begin{equation}
P_{l}(k,1)\simeq\frac{\Lambda_{l}}{\sqrt{4\pi}k^{3/2}}e^{-\frac{l^{2}}{4\lambda^{2}k}},\label{601}
\end{equation}
where $\Lambda_{l}$ is given in (\ref{503d}). An interesting picture
emerges from the above-mentioned results. Downstream sites with $k\gg1$
are mostly empty. However, conditioned on being occupied, their occupation
is typically of the order of $\sqrt{k}$ {[}see Eq. (\ref{305}){]},
and in fact its distribution has the scaling form of Eq. (\ref{601}).\textcolor{red}{{}
}Below, in Section \ref{sec:Homogeneous-ASIPs-Exact}, we derive an
exact expression for this occupation probability which is correct
even for small $k$.

\emph{Inter-particle distance probability}. Another quantity of interest
is the inter-particle distance probability $Q(k,m)$, which is defined
as the conditional probability that the next occupied site after site
$k$ is site $k+m$ given that site $k$ itself is occupied. The scaling
solutions found in Section \ref{sec:Continuum-limit-of} allow us
to calculate $Q(k,m)$. To do so, we first examine the unconditional
probability $\left(1-P_{0}(k,1)\right)Q(k,m)$ that sites $k$ and
$k+m$ are both occupied and the $m-1$ sites in between the two are
empty. This probability is given by
\begin{equation}
\begin{array}{l}
\left(1-P_{0}(k,1)\right)Q(k,m)=P_{0}(k+1,m-1)\\
\\
-\left[P_{0}(k+1,m)-P_{0}(k,m+1)\right]\\
\\
-\left[P_{0}(k,m)-P_{0}(k,m+1)\right]-P_{0}(k,m+1).\\
\text{ }
\end{array}\label{602a}
\end{equation}
The first term in Eq. (\ref{602a}) is the probability that sites
$\{k+1,\cdots,k+m-1\}$ are empty. From this probability one must
subtract: (i) the probability that these sites are empty, site $k$
is occupied and site $k+m$ is empty (the second term, in square brackets);
(ii) the probability that these sites are empty, site $k$ is empty
and site $k+m$ is occupied (the third term, in square brackets);
(iii) the probability that all $m+1$ sites from $k$ to $k+m$ are
empty (the last term). Rearranging and passing, as before, to a continuum
limit yields{\footnotesize{
\begin{equation}
\begin{array}{l}
\left(1-P_{0}(k,1)\right)Q(k,m)=\\
\\
+\left[P_{0}(k,m+1)-2P_{0}(k,m)+P_{0}(k,m-1)\right]\\
\\
-\left[P_{0}(k+1,m)-P_{0}(k,m)-P_{0}(k+1,m-1)+P_{0}(k,m-1)\right]\\
\\
\simeq\left(\frac{\partial^{2}}{\partial m^{2}}-\frac{\partial}{\partial m\partial k}\right)P_{0}(k,m)|_{k,m}\,.\\
\text{ }
\end{array}\label{602a-1}
\end{equation}
}}Substituting Eq. (\ref{502c}) we see that the $\partial^{2}/\partial m^{2}$
term dominates in the large $k$ limit, and we obtain 
\begin{equation}
Q(k,m)\simeq\frac{f''(u)}{k\left(1-P_{0}(k,1)\right)}=\frac{ue^{-u^{2}/4}}{2\sqrt{k}},\label{603}
\end{equation}
where once again $f(u)=\textrm{erfc}(u/2)$ and $u=m/\sqrt{k}$.

\emph{Inter-exit times}. For an ASIP in steady state let the random
variable $T_{k}$ denote the time elapsing between two consecutive
time epochs at which particles exit site $k$. Equation (\ref{603})
allows us to evaluate the typical order of magnitude of $T_{k}$ in
the limit of $k\gg1$. Indeed, given that site $k$ is occupied, it
will take (on average) a single time unit for particles to hop out
of it --- resulting in a first exit event. On the other hand, we know
that $Q(k-m,m)$ is the probability that $k-m$ is the nearest occupied
site in the upstream direction. The average distance to the nearest
occupied site is hence

\begin{equation}
\begin{array}{l}
\sum_{m=1}^{k-1}mQ(k-m,m)\simeq\\
\\
\sqrt{k}\overset{\sqrt{k}}{\underset{0}{\int}}\frac{u^{2}e^{-u^{2}/4\left(1-u/\sqrt{k}\right)}}{2\sqrt{1-u/\sqrt{k}}}du\simeq\sqrt{\pi k}\,.\\
\text{ }
\end{array}\label{603a}
\end{equation}
Thus, $1+\sqrt{\pi k}$ sites on average are to be traversed at an
average `speed' of one site per unit time for the second exit event
to occur. When $k$ is large, $T_{k}$ is clearly dominated by this
traversal time. The error incurred by neglecting the time awaited
till the occurrence of the first exit event is negligible and we may
safely conclude that $\left\langle T_{k}\right\rangle /\sqrt{\pi k}\simeq1$. 

We can further go on and compute the asymptotic distribution of the
inter-exit time. To see how, note that in the limit of $k\gg1$, the
reasoning given above asserts that the probability density of the
random variable $T_{k}$ may be approximated by 
\begin{equation}
P{}_{T_{k}}(t)\simeq\sum_{m=1}^{k-1}\frac{t^{m}e^{-t}}{m!}Q(k-m,m)\label{604}
\end{equation}
where $t^{m}e^{-t}/m!$ is the probability density for the traversal
time of $m+1$ sites. In Appendix \ref{sub:Saddle-point-evaluation 604}
we show that the sum in (\ref{604}) can be evaluated using a saddle
point approximation to yield Eq. (\ref{308}).

\section{\label{sec:Incremental-Load-Analysis}Incremental Load Analysis}

The analysis conducted so far was based on a continuum limit approximation
of Eq. (\ref{409}) at steady state. Using this approach we were able
to analyze homogenous ASIPs and obtain an asymptotic solution for
the probabilities $P_{l}(k,m)$ in the limit $k\gg m,l$. We now set
forth to obtain an \emph{exact} solution for this problem. In order
to demonstrate the general applicability of the approach described
hereinafter we develop it in the context of general ASIPs (not necessarily
homogeneous). Setting off from the stochastic law of motion of the
incremental load, we go on to derive the boundary value problem which
governs its steady state distribution. An algorithm for the solution
of this problem is presented along with iterative schemes for the
computation of occupation probabilities and factorial moments. In
the next section we return to the case of homogeneous ASIPs.

\subsection{\label{sub:Overall-Load}The Incremental Load }

In this section we revisit the notion of incremental load, which generalizes
the notion of overall load. In what follows we consider an infinite
lattice with countably many sites, and analyze the ASIP's incremental
load in detail. We consider the lattice interval starting at site
$k$ and consisting of $m$ sites --- $\{k,k+1,\cdots,k+m-1\}$ ($k,m=1,2,3,\cdots$)
--- and remind the reader that the ASIP's incremental loads $L\left(k,m\right)$
and $L\left(t;k,m\right)$ are given by Eqs. (\ref{309}) and (\ref{401}),
respectively. 

Throughout this section we shall employ the natural boundary conditions
$L\left(t;k,0\right)=0$ and $L\left(k,0\right)=0$. The Probability
Generating Functions (PGFs) of the incremental loads $L\left(k,m\right)$
and $L\left(t;k,m\right)$ are given, respectively, by 
\begin{equation}
G\left(z;k,m\right)=\left\langle z^{L\left(k,m\right)}\right\rangle \,,\label{702a}
\end{equation}
and
\begin{equation}
G\left(t,z;k,m\right)=\left\langle z^{L\left(t;k,m\right)}\right\rangle \,,\label{702b}
\end{equation}
($\left\vert z\right\vert \leq1$). Note that the boundary conditions
$L\left(t;k,0\right)=0$ and $L\left(k,0\right)=0$ imply, respectively,
the following PGF boundary conditions: 
\begin{equation}
G\left(z;k,0\right)=1\text{ \ and \ }G\left(t,z;k,0\right)=1.\label{702c}
\end{equation}

\subsection{\label{sub:The-Case-of k=00003D1}The Case of $k=1$}

In this Subsection we analyze the special case of lattice intervals
initiating at the first lattice site $k=1$. This special case yields
the overall load which was analyzed in \cite{ASIP-1} via the ASIP's
multidimensional PGF. Here we analyze this special case via the method
of incremental loads. This serves to illustrate the method which will
later be used to derive new results for $k>1$. 
\begin{figure}
\begin{centering}
\includegraphics[scale=0.37]{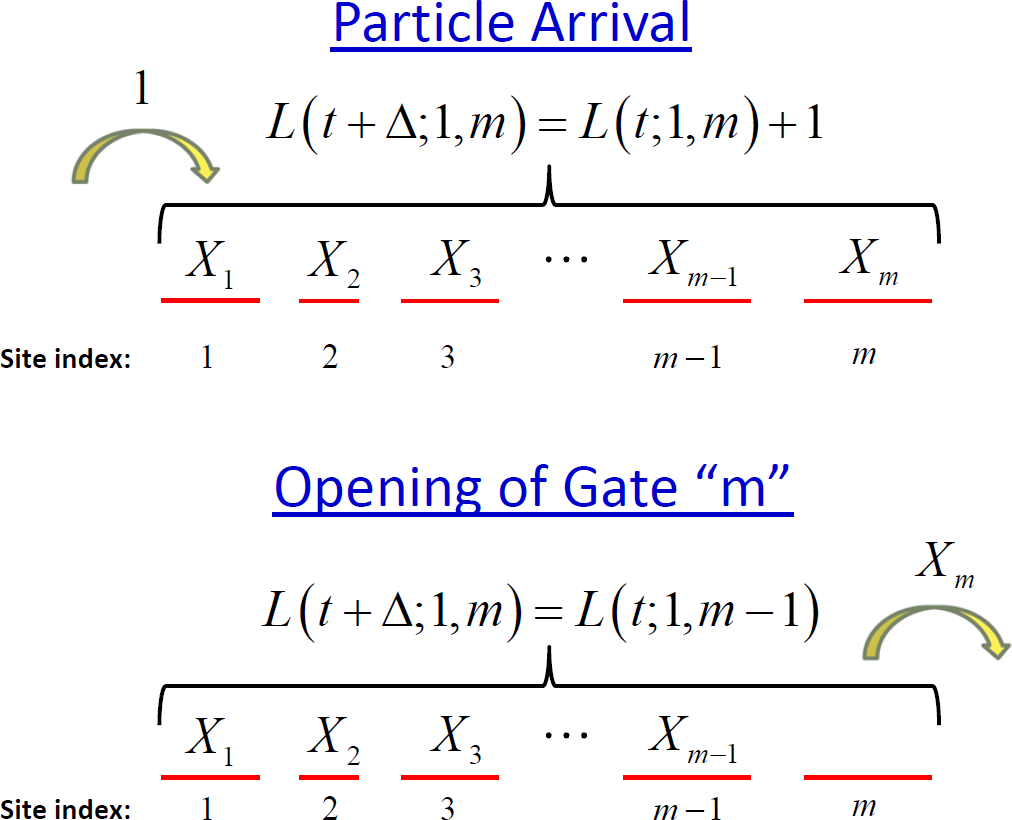}
\par\end{centering}

\caption{(color online). During the time interval $\left(t,t+\triangle\right)$,
the incremental load $L\left(t;1,m\right)$ can change either due
to the arrival of a particle to the first site, or due to the opening
of gate $m$. }
\end{figure}
{\small{ }}{\small \par}

Consider the lattice interval starting at site $1$ and consisting
of $m$ sites, and observe its incremental load at times $t$ and
$t^{\prime}=t+\Delta$ (where $\Delta\rightarrow0$). During the time
interval $\left(t,t^{\prime}\right)$ exactly two events, illustrated
in Figure 3, can change the incremental load. One event is the arrival
of a particle to the lattice --- in which case the arriving particle
enters the first site and hence $L\left(t^{\prime};1,m\right)=L\left(t;1,m\right)+1$;
this event occurs with probability $\lambda\Delta+o(\Delta)$. The
other event is the opening of gate $m$ --- in which case all particles
present in site $m$ transit to site $m+1$ and hence $L\left(t^{\prime};1,m\right)=L\left(t;1,m-1\right)$;
this event occurs with probability $\mu_{m}\Delta+o(\Delta)$. Note
that the boundary condition $L\left(t;1,0\right)=0$ indeed fits in
naturally. If neither of these two events take place --- a scenario
occurring with probability $1-\left(\lambda+\mu_{m}\right)\Delta+o(\Delta)$
--- then the incremental load is left unchanged: $L\left(t^{\prime};1,m\right)=L\left(t;1,m\right)$.
Thus, the stochastic connection between the incremental loads $L\left(t;1,m\right)$
and $L\left(t^{\prime};1,m\right)$ is given by
\begin{equation}
\begin{array}{l}
L\left(t^{\prime};1,m\right)=\\
\text{ }\\
\left\{ \begin{array}{lll}
L\left(t;1,m\right)+1 &  & \text{w.p. \ }\lambda\Delta+o(\Delta)\text{ ,}\\
 & \text{ \ }\\
L\left(t;1,m-1\right) &  & \text{w.p. \ }\mu_{m}\Delta+o(\Delta)\text{ ,}\\
 & \text{ \ }\\
L\left(t;1,m\right) &  & \text{w.p. \ }1-\left(\lambda+\mu_{m}\right)\Delta+o(\Delta)\text{ , }
\end{array}\right.
\end{array}\label{703}
\end{equation}
and we note that ``w.p.'' is used here as a short hand for the term
``with probability''. 

Shifting from the incremental loads $L\left(t;1,m\right)$ and $L\left(t^{\prime};1,m\right)$
to their respective PGFs, Eq. (\ref{703}) yields the following PGF
dynamics 
\begin{equation}
\begin{array}{l}
\frac{\partial}{\partial t}G\left(t,z;1,m\right)=\\
\text{ }\\
\left[\lambda\left(z-1\right)-\mu_{m}\right]G\left(t,z;1,m\right)+\mu_{m}G\left(t,z;1,m-1\right)\text{ .}
\end{array}\label{704}
\end{equation}
The derivation of Eq. (\ref{704}) is given in Appendix \ref{sub:Derivation-of-Eq. 704}.
At steady state the time-dependence vanishes, and the differential
equation (\ref{704}) reduces to the steady-state equation 
\begin{equation}
G\left(z;1,m\right)=\frac{\mu_{m}}{\mu_{m}+\lambda\left(1-z\right)}G\left(z;1,m-1\right)\text{ .}\label{705}
\end{equation}
A straightforward iterative solution of Eq. (\ref{705}), using the
PGF boundary condition $G\left(z;1,0\right)=1$, yields the following
explicit form for the PGF of the incremental load at steady state
\begin{equation}
G\left(z;1,m\right)=\prod_{i=1}^{m}\frac{1}{1+\frac{\lambda}{\mu_{i}}\left(1-z\right)}\,.\text{ }\label{706}
\end{equation}
Note that the terms $\lambda/\mu_{i}$ appearing in Eq. (\ref{706})
are the ratios of the particles' inflow rate to the gates' opening
rates, as well as the mean occupancies at steady state ($\lambda/\mu_{i}=\left\langle X_{i}\right\rangle $)
\cite{ASIP-1}.

Eq. (\ref{706}) has several important implications. Firstly, Eq.
(\ref{706}) implies that at steady state the overall load $L\left(1,1\right)$
of a \emph{single-site} ASIP ($n=1$) follows a \emph{geometric distribution}.
Indeed, setting $m=1$ in Eq. (\ref{706}) yields the PGF of the following
geometric probability distribution: $\Pr\left(L\left(1,1\right)=l\right)=(1-p_{1})^{l}p_{1}$
($l=0,1,2\cdots$), where $p_{1}=\mu_{1}/\left(\mu_{1}+\lambda\right)$.
Secondly, the \emph{product-form }structure of Eq. (\ref{706}) implies
that at steady state the overall load $L\left(1,m\right)$ admits
the stochastic representation
\begin{equation}
L(1,m)=\sum_{i=1}^{m}G_{i}\text{ ,}\label{707}
\end{equation}
where $\left\{ G_{1},\cdots,G_{m}\right\} $ is a sequence of independent
geometrically-distributed random variables: $\Pr\left(G_{i}=l\right)=(1-p_{i})^{l}p_{i}$
($l=0,1,2,\cdots$), with $p_{i}=\mu_{i}/\left(\mu_{i}+\lambda\right)$
($i=1,\cdots,m$). The overall load $L\left(1,m\right)$ is hence
equal, in law, to the sum of the overall loads of $m$ \emph{independent}
\emph{single-site} ASIPs with respective parameters $\left(\lambda,\mu_{1}\right),\cdots,\left(\lambda,\mu_{m}\right)$.
Thus, the distribution of the overall load $L\left(1,m\right)$ is
given by 
\begin{equation}
\begin{array}{l}
P_{l}(1,m)=\Pr\left(L\left(1,m\right)=l\right)=\\
\text{ }\\
\underset{l_{1},\cdots,l_{m}}{\sum}\left(\overset{m}{\underset{i=1}{\prod}}p_{i}(1-p_{i})^{l_{i}}\right)\delta\left(l-\sum_{i}l_{i}\right)\,,
\end{array}\label{708a}
\end{equation}
where the Dirac $\delta$ function guarantees that $\sum_{i}l_{i}=l$.
Thirdly, setting $z=0$ in Eq. (\ref{706}) (or $l=0$ in Eq. (\ref{708a}))
yields the probability that the lattice interval $\left\{ 1,\cdots,m\right\} $
is empty 
\begin{equation}
P_{0}(1,m)=\Pr\left(L\left(1,m\right)=0\right)=\prod_{i=1}^{m}\frac{\mu_{i}}{\mu_{i}+\lambda}\text{ .}\label{708b}
\end{equation}

\subsection{\label{sub:The-Case k>1}The Case $k>1$}

In this Subsection we analyze the general case of lattice intervals
initiating at an arbitrary lattice site $k>1$. While the special
case $k=1$ could be analyzed via the ASIP's joint PGF, an analogous
analysis of the general case $k>1$ via this method is prohibitively
hard. However, as we shall now demonstrate, the analysis of the general
case $k>1$ is well attainable following an approach parallel to the
one applied in the previous Subsection.
\begin{figure}
\begin{centering}
\includegraphics[scale=0.5]{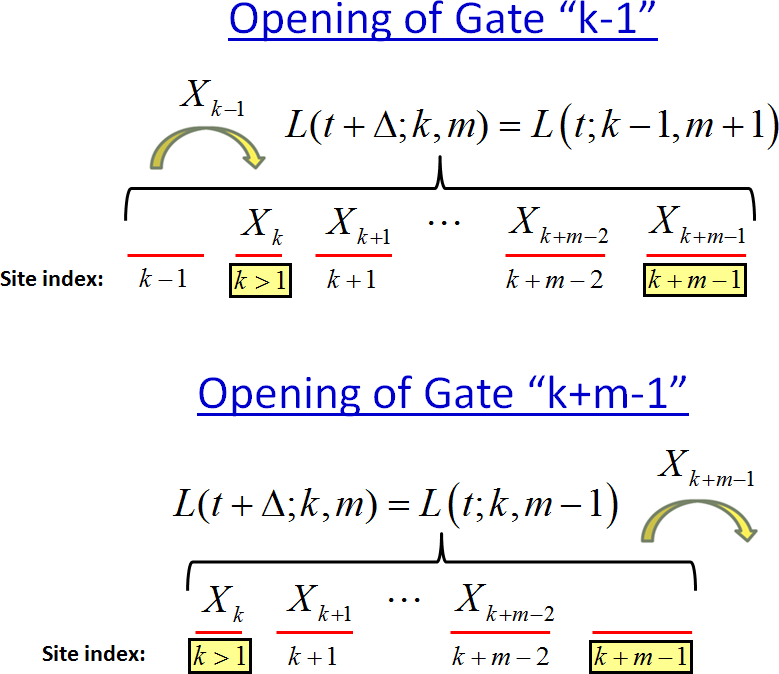}
\par\end{centering}

\caption{(color online). During the time interval $\left(t,t+\triangle\right)$,
the incremental load $L\left(t;k,m\right)$ can change either due
to the opening of gate $k-1$ or due to the opening of gate $k+m-1$. }
\end{figure}
{\small{ }}{\small \par}

Consider the lattice interval starting at site $k$ ($k>1$) and consisting
of $m$ sites, and observe its incremental load at times $t$ and
$t^{\prime}=t+\Delta$ (where $\Delta\rightarrow0$). During the time
interval $\left(t,t^{\prime}\right)$ exactly two events, illustrated
in Figure 4, can change the incremental load. One event is the opening
of gate $k-1$ --- in which case all particles present in site $k-1$
transit to site $k$ and hence $L\left(t^{\prime};k,m\right)=L\left(t;k-1,m+1\right)$;
this event occurs with probability $\mu_{k-1}\Delta+o(\Delta)$. The
other event is the opening of gate $k+m-1$ --- in which case all
particles present in site $k+m-1$ transit to site $k+m$ and hence
$L\left(t^{\prime};k,m\right)=L\left(t;k,m-1\right)$; this event
occurs with probability $\mu_{k+m-1}\Delta+o(\Delta)$. As noted in
Subsection \ref{sub:The-Case-of k=00003D1}, the boundary condition
$L\left(t;k,0\right)=0$ fits in naturally. If neither of these two
events take place --- a scenario occurring with probability $1-\left(\mu_{k-1}+\mu_{k+m-1}\right)\Delta+o(\Delta)$
--- then the incremental load is left unchanged: $L\left(t^{\prime};k,m\right)=L\left(t;k,m\right)$.
Thus, the stochastic connection between the incremental loads $L\left(t;k,m\right)$
and $L\left(t^{\prime};k,m\right)$ is given by

\begin{equation}
\begin{array}{l}
L\left(t^{\prime};k,m\right)=\\
\text{ }\\
\left\{ \begin{array}{ll}
L\left(t;k-1,m+1\right) & \text{w.p. \ }\mu_{k-1}\Delta\text{ ,}\\
\\
L\left(t;k,m-1\right) & \text{w.p. \ }\mu_{k+m-1}\Delta\text{ ,}\\
\\
L\left(t;k,m\right) & \text{w.p. \ }1-\left(\mu_{k-1}+\mu_{k+m-1}\right)\Delta\text{ .}
\end{array}\right.
\end{array}\label{709}
\end{equation}
{\small{ }}{\small \par}

Shifting from the incremental loads $L\left(t;k,m\right)$ and $L\left(t^{\prime};k,m\right)$
to their respective PGFs, Eq. (\ref{709}) yields the following PGF
dynamics
\begin{equation}
\begin{array}{l}
\frac{\partial}{\partial t}G\left(t,z;k,m\right)=-\left(\mu_{k-1}+\mu_{k+m-1}\right)G\left(t,z;k,m\right)\\
\\
\text{ }\\
+\mu_{k-1}G\left(t,z;k-1,m+1\right)+\mu_{k+m-1}G\left(t,z;k,m-1\right)\,\,.
\end{array}\label{710}
\end{equation}
The derivation of Eq. (\ref{710}) is given in Appendix \ref{sub:Derivation-of-Eq. 710}.
At steady state the time-dependence vanishes, and the differential
equation (\ref{710}) reduces to the steady-state equation 
\begin{equation}
\begin{array}{l}
G\left(z;k,m\right)=\frac{\mu_{k+m-1}}{\mu_{k-1}+\mu_{k+m-1}}G\left(z;k,m-1\right)\\
\text{ }\\
+\frac{\mu_{k-1}}{\mu_{k-1}+\mu_{k+m-1}}G\left(z;k-1,m+1\right)\text{ .}
\end{array}\label{711}
\end{equation}
For any fixed $z$, Eq. (\ref{711}) defines a two-dimensional boundary
value problem for $G\left(z;k,m\right)$. The problem and an algorithm
for its solution are illustrated in Figure 5. 
\begin{figure}
\begin{centering}
\includegraphics[scale=0.5]{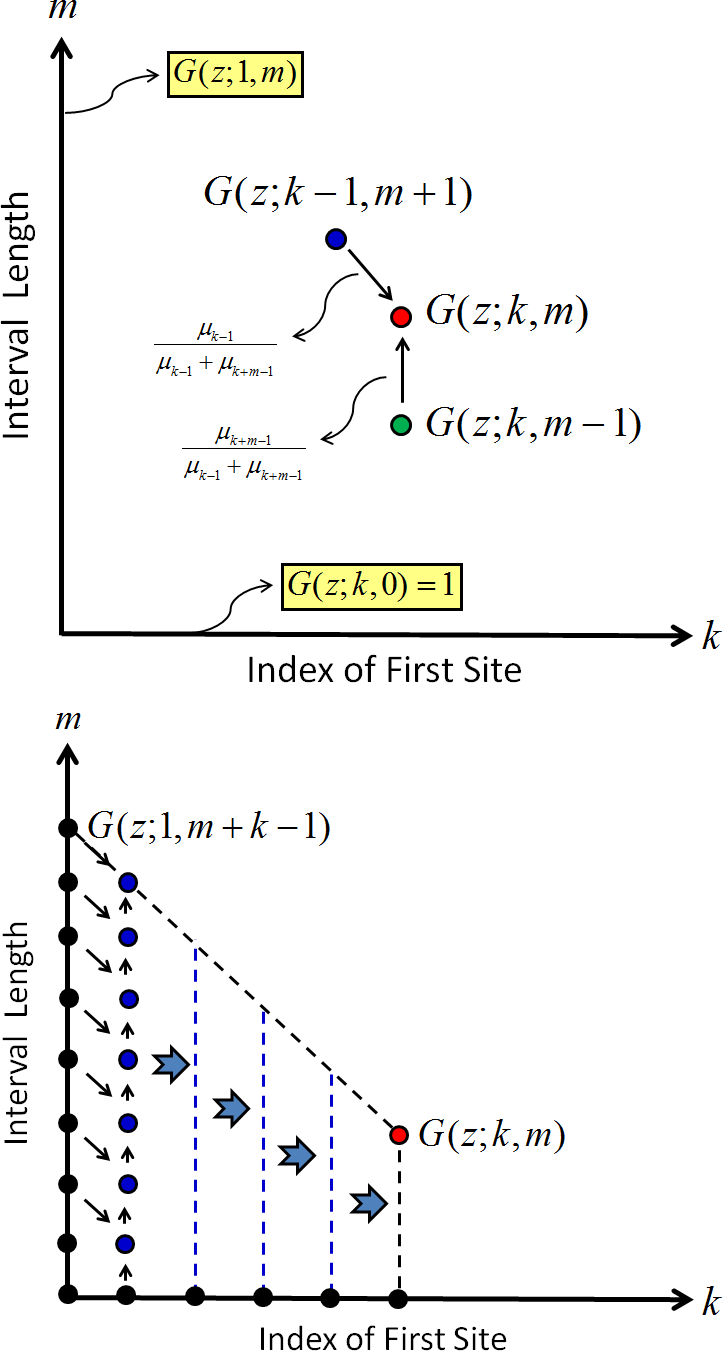}
\par\end{centering}

\caption{(color online). Top Panel: Equation (\ref{711}) defines a boundary
value problem for $G\left(z;k,m\right)$. The PGF $G\left(z;k,m\right)$
is determined by a weighted average of its southern and northwestern
neighbors in the positive quadrant of the $(k,m)$ plane. The boundary
PGFs $G\left(z;k,0\right)$ and $G\left(z;1,m\right)$ are given by
Eqs. (\ref{702c}) and (\ref{706}) respectively. Bottom Panel: A
three step algorithm can be used in order to solve the boundary value
problem for the PGF $G\left(z;k,m\right)$: (i) start at the left
boundary and solve for the column that stands to its right; (ii) treat
the newly solved column as the new left boundary and iterate; (iii)
stop at the $k^{th}$ column and obtain the desired solution.   }
\end{figure}

Equation (\ref{711}) can also be used to establish an explicit iterative
scheme for the computation of the PGF $G\left(z;k,m\right)$ in terms
of the PGFs $\left\{ G\left(z,k-1,i\right)\right\} _{i=2,...,m+1}$.
Specifically:
\begin{equation}
\begin{array}{l}
G\left(z;k,m\right)=\Pi\left(k,m\right)\\
\text{ }\\
+\Pi\left(k,m\right)\sum_{i=1}^{m}\frac{\mu_{k-1}}{\mu_{k-1}+\mu_{k+i-1}}\frac{G\left(z;k-1,i+1\right)}{\Pi\left(k,i\right)}\text{ ,}
\end{array}\label{712}
\end{equation}
where
\begin{equation}
\Pi\left(k,m\right)=\prod_{j=1}^{m}\frac{\mu_{k+j-1}}{\mu_{k-1}+\mu_{k+j-1}}\text{ ,}\label{713}
\end{equation}
and where the boundary condition $G\left(z;1,m\right)$ is given by
Eq. (\ref{706}). The derivation of Eq. (\ref{712}) is given in Appendix
\ref{sub:Derivation-of-Eq. 712}.

\subsection{\label{sub:Occupation-Probabilities-and-Factorial-Moments1}Occupation
Probabilities and Factorial Moments}

Based on the incremental-load results established hitherto, in this
Subsection we derive recursive equations for the occupation probabilities
and the factorial moments of the incremental loads. We begin with
the occupation probabilities, and then turn to the factorial moments. 

In terms of the PGF $G\left(z;k,m\right)$ the steady state probability
of finding exactly $l$ particles ($l=0,1,2,\cdots)$ in the interval
$\left\{ k,k+1,\cdots,k+m-1\right\} $ is given by

\begin{equation}
P_{l}\left(k,m\right)=\frac{1}{l!}\left.\frac{d^{l}}{dz^{l}}G\left(z;k,m\right)\right\vert _{z=0}\,,\label{715}
\end{equation}
with $P_{0}\left(k,m\right)=G\left(0;k,m\right)$. Taking the $l^{th}$
derivative of Eq. (\ref{712}) with respect to the variable $z$,
setting $z=0$ and dividing by $l!$, Eq. (\ref{715}) yields the
following recursion for the occupation probabilities 
\begin{equation}
\begin{array}{l}
\mathrm{P_{\mathit{l}}(\mathit{k,m})}=\Pi\left(k,m\right)\delta_{l,0}\\
\\
\text{ }\\
+\Pi\left(k,m\right)\sum_{i=1}^{m}\frac{\mu_{k-1}}{\mu_{k-1}+\mu_{k+i-1}}\frac{\mathrm{P_{\mathit{l}}(\mathit{k-1,i+1})}}{\Pi\left(k,i\right)}\text{\thinspace.}
\end{array}\label{716}
\end{equation}
Equation (\ref{716}), together with the boundary condition in Eq.
(\ref{708a}), establishes an explicit iterative scheme for the computation
of the occupation probabilities $\mathrm{P_{\mathit{l}}(\mathit{k,m})}$
in terms of the occupation probabilities $\left\{ \mathrm{P_{\mathit{l}}(\mathit{k-\mathrm{1},\mathrm{i}})}\right\} _{i=2,\cdots,m+1}$.

Analogously, one can further derive recursive equations for the factorial
moments of the incremental load $L\left(k,m\right)$. In terms of
the PGF $G\left(z;k,m\right)$, the factorial moments $M_{l}\left(k,m\right)$
$(l=1,2,3,\cdots)$ are given by
\begin{equation}
M_{l}\left(k,m\right)\equiv\left\langle \overset{l-1}{\underset{i=0}{\prod}}\left(L\left(k,m\right)-i\right)\right\rangle =\left.\frac{d^{l}}{dz^{l}}G\left(z;k,m\right)\right\vert _{z=1}\,.\label{717}
\end{equation}
Hence, taking the $l^{th}$ derivative of Eq. (\ref{712}) with respect
to the variable $z$ and setting $z=1$, Eq. (\ref{717}) yields the
following recursive equation for the factorial moments 
\begin{equation}
\begin{array}{l}
M_{l}\left(k,m\right)=\\
\text{ }\\
\Pi\left(k,m\right)\sum_{i=1}^{m}\frac{\mu_{k-1}}{\mu_{k-1}+\mu_{k+i-1}}\frac{M_{l}\left(k-1,i+1\right)}{\Pi\left(k,i\right)}\text{ .}
\end{array}\label{720}
\end{equation}
Equation (\ref{720}), together with the boundary condition
\begin{equation}
M_{l}\left(1,m\right)=\left.\frac{d^{l}}{dz^{l}}\prod_{i=1}^{m}\frac{1}{1+\frac{\lambda}{\mu_{i}}\left(1-z\right)}\right\vert _{z=1}\text{ }\label{721}
\end{equation}
establishes an explicit iterative scheme for the computation of the
factorial moments $M_{l}\left(k,m\right)$ in terms of the factorial
moments $\left\{ M_{l}\left(k-1,i\right)\right\} _{i=2,\cdots,m+1}$.

\section{\label{sec:Homogeneous-ASIPs-Exact}Incremental Load: Exact Results }

In this section we return to the analysis of homogeneous ASIPs and
provide exact results for the occupation probabilities and the factorial
moments of the incremental load $L(k,m)$. These results are given
in terms of the Catalan trapezoid --- a generalization of the well
known Catalan numbers which appear in many combinatorial settings
\cite{Catalan}. We start in Subsection \ref{sub:The-Catalan-Numbers}
with a prelude on the Catalan numbers and their extensions, as these
numbers will prove instrumental in our analysis. Then, in Subsection
\ref{sub:Occupation-Probabilities-and-exact} we present exact results
for incremental load probabilities and factorial moments. We conclude
in Subsection \ref{sub:The-probability-generating} in which we provide
a derivation of the results presented in \ref{sub:Occupation-Probabilities-and-exact}
along with exact results for the probability generating function of
the incremental load.

\subsection{\label{sub:The-Catalan-Numbers}The Catalan Numbers }

Named after the French-Belgian mathematician Eugène Charles Catalan,
these numbers arise in various problems in combinatorics. For concreteness
we shall henceforth address the Catalan numbers in the following context.
Consider a string of numbers composed out of $n$ $(+1)$'s and $n$
$(-1)$'s, arranged in a row from left to right, such that the sum
over every initial substring is non-negative. What is the total number
of different such strings? The solution to this combinatorial problem
is given by the $n^{th}$ Catalan number \cite{Catalan}: 
\begin{equation}
C(n)=\left(\begin{array}{c}
2n\\
n
\end{array}\right)-\left(\begin{array}{c}
2n\\
n-1
\end{array}\right)\label{801}
\end{equation}
$(n=1,2,3,...)$, with $C(0)=1$ by definition. Specifically, the
first Catalan numbers are given by $\{1,1,2,5,14,42,132,429,...\}$.

One can generalize the combinatorial problem mentioned above by considering
a string of $n$ $(+1)$'s and $k$ $(-1)$'s. In this case, the number
of different strings for which the sum over every initial substring
is non-negative is given by 
\begin{equation}
C(n,k)=\begin{cases}
1 & \,\,\, k=0\\
\\
\left(\begin{array}{c}
n+k\\
k
\end{array}\right)-\left(\begin{array}{c}
n+k\\
k-1
\end{array}\right) & \,\,\,1\leq k\leq n\\
\\
0 & \,\,\, k>n
\end{cases}\label{802}
\end{equation}
$(n=0,1,2,\cdots;k=0,1,2,\cdots)$. 

The numbers $C(n,k)$ are referred to --- in combinatorial mathematics
--- as the entries of \emph{Catalan's triangle} \cite{Catalan,Catalan's_triangle1,Catalan's_triangle2,Catalan's_triangle3}.
These numbers facilitate the solution to Bertrand's ballot problem:
``In an election where candidate A receives $n$ votes and candidate
$B$ receives $k$ votes, what is the probability that $A$ will not
trail behind $B$ throughout the entire count of votes?''. Indeed,
the answer to Bertrand's problem is given by the ratio $C(n,k)/\left(\begin{array}{c}
n+k\\
k
\end{array}\right)$. 

Catalan's triangle, illustrated in Table I, has the following iterative
construction. By definition, all entries that are positioned on the
left boundary of the triangle $(k=0)$ are given by the boundary condition
$C(n,0)=1$; in Table I, these entries are highlighted in bold. Entries
positioned to the right of the main diagonal $k=n$ are zero; in Table
I, these entries are indicated by empty squares. All the other entries
of Catalan's triangle follow the recursion 
\begin{equation}
C(n,k)=C(n-1,k)+C(n,k-1)\,,\label{803}
\end{equation}
i.e., each entry is a sum of the entry above it and the entry to its
left; in Table I, a specific example, $20+7=27$, is highlighted in
magenta. Entries on the diagonal of Catalan's triangle $(k=n)$ are
the Catalan numbers: $C(n,n)=C(n)$; in Table I these entries are
highlighted in blue. 
\begin{table}
\begin{centering}
\begin{tabular}{|c||c|c|c|c|c|c|c|c|}
\hline 
$n/k$ & 0 & 1 & 2 & 3 & 4 & 5 & 6 & 7\tabularnewline
\hline 
\hline 
0 & \textbf{\textcolor{blue}{1}} &  &  &  &  &  &  & \tabularnewline
\hline 
1 & \textbf{\textcolor{green}{1}} & \textcolor{blue}{1} &  &  &  &  &  & \tabularnewline
\hline 
2 & \textbf{\textcolor{red}{1}} & \textcolor{green}{2} & \textcolor{blue}{2} &  &  &  &  & \tabularnewline
\hline 
3 & \textbf{1} & \textcolor{red}{3} & \textcolor{green}{5} & \textcolor{blue}{5} &  &  &  & \tabularnewline
\hline 
4 & \textbf{1} & 4 & \textcolor{red}{9} & \textcolor{green}{14} & \textcolor{blue}{14} &  &  & \tabularnewline
\hline 
5 & \textbf{1} & 5 & 14 & \textcolor{red}{28} & \textcolor{green}{42} & \textcolor{blue}{42} &  & \tabularnewline
\hline 
6 & \textbf{1} & 6 & \textcolor{magenta}{20} & 48 & \textcolor{red}{90} & \textcolor{green}{132} & \textcolor{blue}{132} & \tabularnewline
\hline 
7 & \textbf{1} & \textcolor{magenta}{7} & \textcolor{magenta}{27} & 75 & 165 & \textcolor{red}{297} & \textcolor{green}{429} & \textcolor{blue}{429}\tabularnewline
\hline 
\end{tabular}
\par\end{centering}

\caption{(color online). Some entries of Catalan's triangle. Entries on the
left boundary are highlighted in bold. Null entries positioned to
the right of the diagonal $k=n$ are indicated by empty squares. All
other entries follow the recursive rule given in Eq. (\ref{803}).
A specific example, $20+7=27$, is highlighted in magenta. The entries
on the diagonal of Catalan's triangle, highlighted in blue, are the
Catalan numbers. The second and third diagonals, highlighted in green
and red respectively, coincide with the main diagonals of Catalan's
trapezoids of order $m=2$ and $m=3$.}
\end{table}

The combinatorial meaning of Eq. (\ref{803}) and its validity for
$1\leq k\leq n$ become immediately clear after conducting a binary
partition of all admissible strings according to their last digit
$+1$ or $-1$. Indeed, since $k\leq n$ the sum over a string of
$n$ $(+1)$'s and $k$ $(-1)$'s is non-negative. Moreover, if the
string ends with $+1$ there are exactly $C(n-1,k)$ ways to choose
the order of the first $n-1$ $(+1)$'s and $k$ $(-1)$'s such that
the sum over every initial substring is non negative. Similarly, if
the string ends with a $-1$ there are exactly $C(n,k-1)$ ways to
choose the order of the first $n$ $(+1)$'s and $k-1$ $(-1)$'s
such that the sum over every initial substring is non negative. Thus,
Eq. (\ref{803}) readily follows. 

Further generalizing the combinatorial problem discussed so far we
now consider the number of different strings of $n$ $(+1)$'s and
$k$ $(-1)$'s for which the sum over every initial substring is larger
than, or equal to, a threshold level $1-m$ $(m=1,2,3,\cdots)$. In
\cite{Catalan Trapezoid} it is shown that this number is given by
{\small{
\begin{equation}
C_{m}(n,k)=\begin{cases}
\left(\begin{array}{c}
n+k\\
k
\end{array}\right) & \,\,\,0\leq k<m\\
\\
\left(\begin{array}{c}
n+k\\
k
\end{array}\right)-\left(\begin{array}{c}
n+k\\
k-m
\end{array}\right) & \,\,\, m\leq k\leq n+m-1\\
\\
0 & \,\,\, k>n+m-1
\end{cases}\label{804}
\end{equation}
}}$(n=0,1,2,\cdots;k=0,1,2,\cdots;m=1,2,3,\cdots)$. Note that $C_{1}(n,k)=C(n,k)$
by definition. Indeed, setting $m=1$ in Eq. (\ref{804}) yields Eq.
(\ref{802}). More generally it can be said that the numbers appearing
in Eq. (\ref{804}) generalize Catalan's triangle to form a countable
family of trapezoid arrays. Fixing the value of the index $m$, we
henceforth refer to $C_{m}(n,k)$ as the entries of the \emph{Catalan’s
trapezoid} of order $m$. Catalan's trapezoid of order $m=2$ and
of order $m=3$ are given in Table II. 

The iterative construction Catalan’s trapezoids is similar to that
of Catalan’s triangle. All elements on the left boundary $(k=0)$
of the trapezoid are given by the boundary condition $C_{m}(n,0)=1$,
all elements on the upper boundary of the trapezoid $(n=0;0\leq k\leq m-1)$
are given by the boundary condition $C_{m}(0,k)=1$ and all elements
positioned to the right of the diagonal $k=n+m-1$ are set to be zero.
The rest of the elements in the trapezoid follow a recursive rule
similar to the one given in Eq. (\ref{803}), albeit replacing the
numbers $C(n,k)$ by the numbers $C_{m}(n,k)$: 
\begin{equation}
C_{m}(n,k)=C_{m}(n-1,k)+C_{m}(n,k-1)\,,\label{805a}
\end{equation}
i.e., each entry is a sum of the entry above it and the entry to its
left. Finally we note an important identity that will come in handy
later on:
\begin{equation}
C_{m}(n,n+m-1)=C_{1}(n+m-1,n)\,.\label{805b}
\end{equation}
That is, the main diagonal of Catlan's trapezoid of order $m$ coincides
with the $m^{th}$ diagonal of Catlan's triangle. This identity is
easily verified by use of Eqs. (\ref{802}) and (\ref{804}). 
\begin{table}[t]
\begin{centering}
\begin{tabular}{|c||c|c|c|c|c|c|c|c|c|}
\hline 
$n/k$ & 0 & 1 & 2 & 3 & 4 & 5 & 6 & 7 & 8\tabularnewline
\hline 
\hline 
0 & \textbf{1} & \textbf{\textcolor{green}{1}} &  &  &  &  &  &  & \tabularnewline
\hline 
1 & \textbf{1} & 2 & \textcolor{green}{2} &  &  &  &  &  & \tabularnewline
\hline 
2 & \textbf{1} & 3 & 5 & \textcolor{green}{5} &  &  &  &  & \tabularnewline
\hline 
3 & \textbf{1} & 4 & 9 & 14 & \textcolor{green}{14} &  &  &  & \tabularnewline
\hline 
4 & \textbf{1} & 5 & 14 & 28 & 42 & \textcolor{green}{42} &  &  & \tabularnewline
\hline 
5 & \textbf{1} & 6 & 20 & 48 & 90 & 132 & \textcolor{green}{132} &  & \tabularnewline
\hline 
6 & \textbf{1} & 7 & 27 & 75 & 165 & 297 & \textcolor{magenta}{429} & \textcolor{green}{429} & \tabularnewline
\hline 
7 & \textbf{1} & 8 & 35 & 110 & 275 & \textcolor{magenta}{572} & \textcolor{magenta}{1001} & 1430 & \textcolor{green}{1430}\tabularnewline
\hline 
\end{tabular}~~\\
~\\
~\\

\par\end{centering}

\begin{centering}
\begin{tabular}{|c||c|c|c|c|c|c|c|c|c|c|}
\hline 
$n/k$ & 0 & 1 & 2 & 3 & 4 & 5 & 6 & 7 & 8 & 9\tabularnewline
\hline 
\hline 
0 & \textbf{1} & \textbf{1} & \textbf{\textcolor{red}{1}} &  &  &  &  &  &  & \tabularnewline
\hline 
1 & \textbf{1} & 2 & 3 & \textcolor{red}{3} &  &  &  &  &  & \tabularnewline
\hline 
2 & \textbf{1} & 3 & 6 & 9 & \textcolor{red}{9} &  &  &  &  & \tabularnewline
\hline 
3 & \textbf{1} & 4 & 10 & 19 & 28 & \textcolor{red}{28} &  &  &  & \tabularnewline
\hline 
4 & \textbf{1} & 5 & 15 & 34 & 62 & 90 & \textcolor{red}{90} &  &  & \tabularnewline
\hline 
5 & \textbf{1} & 6 & 21 & 55 & \textcolor{magenta}{117} & 207 & 297 & \textcolor{red}{297} &  & \tabularnewline
\hline 
6 & \textbf{1} & 7 & 28 & \textcolor{magenta}{83} & \textcolor{magenta}{200} & 407 & 704 & 1001 & \textcolor{red}{1001} & \tabularnewline
\hline 
7 & \textbf{1} & 8 & 36 & 119 & 319 & 726 & 1430 & 2431 & 3432 & \textcolor{red}{3432}\tabularnewline
\hline 
\end{tabular}
\par\end{centering}

\caption{(color online). Some entries of Catalan's trapezoid of order $m=2$
(top) and $m=3$ (bottom). Entries on the left and upper boundaries
are highlighted in bold. Null entries positioned to the right of the
diagonal $k=n+m-1$ are indicated by empty squares. All other entries
follow the recursive rule given in Eq. (\ref{805a}). Two specific
examples, $429+572=1001$ and $117+83=200$, are highlighted in magenta.
The main diagonals of Catalan's trapezoids of order $m=2$ and $m=3$,
highlighted in green and red respectively, coincide with the second
and third diagonals of Catlan's triangle (highlighted, in Table I,
in green and red respectively). }
\end{table}

\subsection{\label{sub:Occupation-Probabilities-and-exact}Occupation Probabilities
and Factorial Moments}

We are now in a position to present exact steady-state results for
both the occupation probabilities and the factorial moments of the
incremental load in the homogenous ASIP. In what follows we return
to the convention by which $\mu=1$ and $\lambda$ is measured in
units of the gate opening rate. The results presented herein will
be expressed in terms of the entries of Catalan's trapezoids $C_{m}(n,k)$.
Detailed proofs are given in the following Subsection. 

We start with the incremental load $L\left(1,m\right)$. Substituting
$p_{i}\rightarrow1/(1+\lambda)$ in Eq. (\ref{708a}) we obtain the
probabilities $P_{l}(1,m)$ given by Eq. (\ref{501c}). Similarly,
substituting $\lambda/\mu_{i}\rightarrow\lambda$ in Eq. (\ref{721})
we obtain the corresponding factorial moments 
\begin{equation}
M_{l}\left(1,m\right)=\frac{(m+l-1)!}{(m-1)!}\lambda^{l}\,.\text{ }\label{807}
\end{equation}

We now turn to the incremental load $L(k,m)$, with $k>1$. In what
follows we show that the occupation probabilities $P_{l}\left(k,m\right)$
($l=0,1,2,\cdots)$ are given by 
\begin{equation}
\begin{array}{l}
\mathrm{P_{\mathit{l}}(\mathit{k,m})}=\delta_{l,0}\overset{k}{\underset{j=2}{\sum}}\frac{C_{1}(k+m-j-1,k-j)}{2^{2k+m-2j}}\\
\\
\text{ }\\
+\sum_{j=2}^{k+m-1}\frac{C_{m}(k-2,m+k-1-j)P_{l}(1,j)}{2^{2k+m-2-j}}\,.
\end{array}\label{808a}
\end{equation}
and that the factorial moments $M_{l}\left(k,m\right)$ ($l=1,2,\cdots)$
are given by
\begin{equation}
\begin{array}{l}
M_{l}\left(k,m\right)=\sum_{j=2}^{k+m-1}\frac{C_{m}(k-2,m+k-1-j)M_{l}\left(1,j\right)}{2^{2k+m-2-j}}\,.\\
\text{ }
\end{array}\label{809a}
\end{equation}

We note that the sums in Eqs. (\ref{808a}--\ref{809a}) contain a
finite number of explicitly known summands and can thus be used for
exact and efficient calculation of $P_{l}\left(k,m\right)$ and $M_{l}\left(k,m\right)$.
Moreover, in the case of single-site lattice intervals ($m=1)$ the
sums in Eqs. (\ref{808a}--\ref{809a}) can be computed (given Eqs.
(\ref{501c}) and (\ref{807}), and by use of any standard computer
algebra software) to be expressed in terms of standard functions.
Specifically, the probability distribution and the factorial moments
of the random variable $X_{k}$ are given by {\small{
\begin{equation}
\begin{array}{l}
\mathrm{P_{\mathit{l}}(\mathit{k,\mathrm{1}})}=\delta_{l,0}\left(1-\frac{\Gamma(k-1/2)}{\sqrt{\pi}\Gamma(k)}\right)\\
\\
\text{ }\\
+\frac{(1+l)\Gamma(k-3/2)\lambda^{l}}{2\sqrt{\pi}\Gamma(k)(1+\lambda)^{2+l}}\times{}_{2}F_{1}\left(2-k,2+l,4-2k;\frac{2}{1+\lambda}\right)\,,
\end{array}\label{808b}
\end{equation}
}}and
\begin{equation}
\begin{array}{l}
M_{l}\left(k,1\right)=\frac{2^{l}\lambda^{l}\Gamma(1+l/2)\Gamma(k+l/2-1/2)}{\sqrt{\pi}\Gamma(k)}\,,\\
\text{ }
\end{array}\label{809b}
\end{equation}
where $\Gamma(x)$ and $_{2}F_{1}\left(a,b,c;x\right)$ are the Gamma
function and hypergeometric function, respectively. For large $k$,
an asymptotic analysis of the exact expressions (\ref{808a}) and
(\ref{808b}), yields the asymptotic results of Section \ref{sec:Continuum-limit-of}.
The details of this asymptotic analysis are sketched in Appendix \ref{sec:Asymptotic-analysis}.

\subsection{The probability generating function\label{sub:The-probability-generating}}

In this Subsection we derive an expression for the probability generating
function $G\left(z;k,m\right)$ and prove the validity of Eqs. (\ref{808a})
and (\ref{809a}). Substituting $\lambda/\mu_{i}\rightarrow\lambda$
in Eq. (\ref{706}) we see that the probability generating function
of the incremental load $L\left(1,m\right)$ is given by 
\begin{equation}
G\left(z;1,m\right)=\left(\frac{1}{1+\lambda\left(1-z\right)}\right)^{m}\,.\text{ }\label{810}
\end{equation}
We now turn to derive an expression for $G\left(z;k,m\right)$ in
the case of $k>1$. Our derivation is based on an insightful probabilistic
interpenetration of the boundary value problem that appears in Eq.
(\ref{711}). An alternative derivation which is algebraic in nature
is given in Appendix \ref{sub:Derivation-of-Eq. 811}. 

The first step in our derivation is to note that Eq. (\ref{711})
is linear with respect to the PGFs that compose it. It follows that
$G\left(z;k,m\right)$ can be expressed as a weighted sum over known
boundary PGFs of the type $G\left(z;1,m\right)$ and $G\left(z;k,0\right)$.
Iterating Eq. (\ref{711}) in an attempt to find the contribution
of a specific boundary PGF to the unknown PGF $G\left(z;k,m\right)$,
we consider a path in the first quadrant of the $(k,m)$ plane that:
(i) is composed out of steps in the south ($\downarrow$) and northwest
($\nwarrow$) directions only; (ii) connects the point $(k,m)$ with
a specific boundary point $(k',m')$ whose position is associated
with the last two arguments of the boundary PGF whose contribution
we are trying to assess; (iii) does not pass through any other boundary
point. A path that complies with the above-mentioned conditions will
henceforth be named a \emph{legitimate path}. 

The number of northwest steps in a legitimate path is given by $k-k'$,
the number of south steps is given by $k-k'+m-m'$ and the total number
of steps is given by $2k-2k'+m-m'$. Since we are dealing with a homogeneous
ASIP, Eq. (\ref{711}) asserts that each step in the path contributes
a multiplicative factor of exactly $1/2$. The contribution due to
a single legitimate path connecting the points $(k,m)$ and $(k',m')$
is hence $\left(1/2\right)^{2k-2k'+m-m'}G\left(z;k',m'\right)$. Taking
into account all possible legitimate paths and summing over all boundary
points we have
\begin{equation}
\begin{array}{l}
G\left(z;k,m\right)=\underset{(k',m')\in boundary}{\sum}\left[\#(k,m,k',m')\right.\\
\text{ }\\
\left.\left(1/2\right)^{2k-2k'+m-m'}G\left(z;k',m'\right)\right]\,,
\end{array}\label{811}
\end{equation}
where $\#(k,m,k',m')$ is the number of legitimate paths that start
at $(k,m)$ and end at $(k',m')$. The idea behind Eq. (\ref{811})
is illustrated in Figure 6. 
\begin{figure}
\begin{centering}
\includegraphics[scale=0.45]{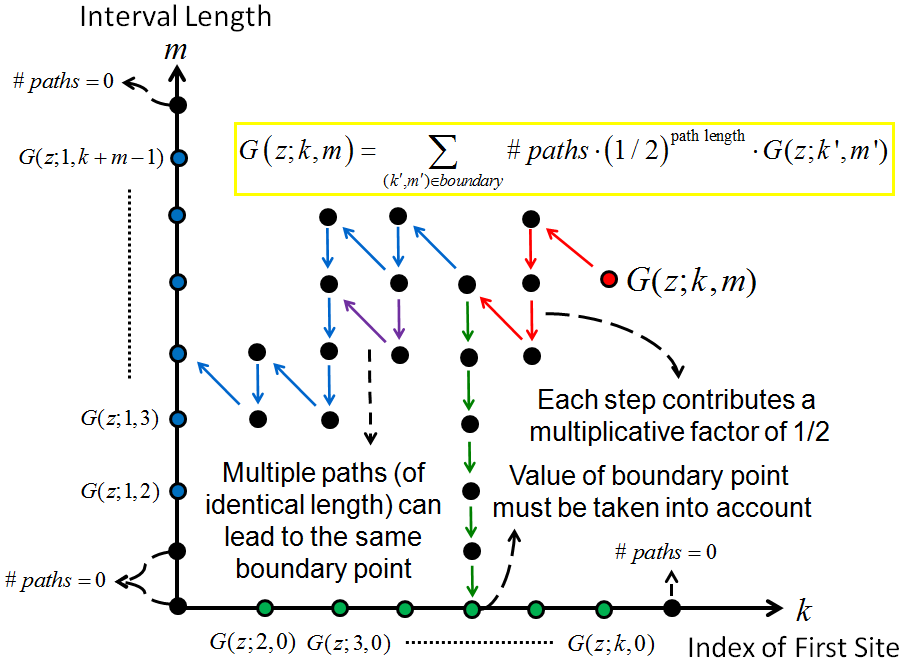}
\par\end{centering}

\caption{(color online). Expressing $G\left(z;k,m\right)$ as a weighted sum
over known boundary functions. All boundary functions must be properly
weighted and taken into account. The weight of each boundary function
is given by the number of legitimate paths leading to it, multiplied
by $1/2$ raised to the power of the path length. Paths are made out
of south ($\downarrow$) and northwest ($\nwarrow$) steps only and
must not pass through another boundary function except the one lying
at the end of the path. Some boundary functions can be reached via
several different paths while others cannot be reached at all (the
latter are discarded in the computation of the sum). }
\end{figure}

In order to proceed we consider a random walker that chooses, with
equal probability at each step, between a south ($\downarrow$) and
northwest step ($\nwarrow$). Assume that the random walker starts
its walk at the point $(k,m)$ and let $P_{\text{hit}}^{k,m}(k',m')$
be the probability that the random walker hits the boundary point
$(k',m')$ before it hits any other boundary point. From this definition
it readily follows that 
\begin{equation}
P_{\text{hit}}^{k,m}(k',m')=\#(k,m,k',m')\cdot\left(1/2\right)^{2k-2k'+m-m'}\,.\label{812}
\end{equation}
We will now show that 
\begin{equation}
P_{\text{hit}}^{k,m}(j,0)=\left(\frac{1}{2}\right)^{2k+m-2j}C_{1}(k+m-j-1,k-j)\label{813}
\end{equation}
($m=1,2,\cdots$;$k=2,3,\cdots$;$j=2,3,\cdots,k$), and that 
\begin{equation}
P_{\text{hit}}^{k,m}(1,j)=\left(\frac{1}{2}\right)^{2k+m-2-j}C_{m}(k-2,m+k-1-j)\text{ }\label{814}
\end{equation}
($m=1,2,\cdots$;$k=2,3,\cdots$;$j=2,3,\cdots,k+m-1$). 

In every legitimate path connecting the point $(k,m)$ with the point
$(j,0)$ ($j=2,3,\cdots,k$) the last step is always directed to the
south. The remaining steps --- $k-j$ northwest and $k-j+m-1$ south
--- must be ordered into a path that connects the point $(k,m)$ to
the point $(j,1)$ without hitting the south boundary first. Similarly,
in every legitimate path connecting the point $(k,m)$ with the point
$(1,j)$ ($j=2,3,\cdots,k+m-1$) the last step is always directed
to the northwest. The remaining steps --- $k-2$ northwest and $k-1+m-j$
south --- must be ordered into a path that connects the point $(k,m)$
to the point $(2,j-1)$ without hitting the south boundary first.
Recalling the combinatorial interpretation of $C_{m}(n,k)$ one can
easily convince himself that $\#(k,m,j,0)=C_{m}(k-j,k-j+m-1)$ and
$\#(k,m,1,j)=C_{m}(k-2,k-1+m-j)$. Equation (\ref{814}) now follows
immediately from Eq. (\ref{812}). Equation (\ref{813}) follows from
Eq. (\ref{812}) by use of the ``diagonal identity'' $C_{m}(k-j,k+m-j-1)=C_{1}(k+m-j-1,k-j)$.

The PGF of the incremental load $L(k,m)$ can now be obtained by substituting
Eq. (\ref{812}) into Eq. (\ref{811}), omitting terms for which $\#(k,m,k',m')=0$,
and utilizing Eqs. (\ref{813})-(\ref{814}) to get 

\begin{equation}
\begin{array}{l}
G\left(z;k,m\right)=\overset{k}{\underset{j=2}{\sum}}G\left(z;j,0\right)\cdot P_{\text{hit}}^{k,m}(j,0)\\
\text{ }\\
+\sum_{j=2}^{k+m-1}G\left(z;1,j\right)\cdot P_{\text{hit}}^{k,m}(1,j)\,.\text{ }
\end{array}\label{815}
\end{equation}
Taking the $l^{th}$ derivative of Eq. (\ref{815}) with respect to
the variable $z$ and setting $z=1$, Eq. (\ref{809a}) follows by
use of Eq. (\ref{717}) and the fact that $G\left(z;j,0\right)=P_{0}(j,0)=1$.
Substituting Eqs. (\ref{813})-(\ref{814}) into Eq. (\ref{815})
we conclude that {\footnotesize{
\begin{equation}
\begin{array}{l}
G\left(z;k,m\right)=\overset{k}{\underset{j=2}{\sum}}G\left(z;j,0\right)\left(\frac{1}{2}\right)^{2k+m-2j}C_{1}(k+m-j-1,k-j)\\
\\
\text{ }\\
+\sum_{j=2}^{k+m-1}G\left(z;1,j\right)\left(\frac{1}{2}\right)^{2k+m-2-j}C_{m}(k-2,m+k-1-j)\,.
\end{array}\label{816}
\end{equation}
}}The occupation probabilities in Eq. (\ref{808a}) can then be read
off from Eq. (\ref{816}) after substituting $G\left(z;j,0\right)=1$
and $G\left(z;1,j\right)=\overset{\infty}{\underset{l=0}{\sum}}P_{l}(1,j)z^{l}$~.

\section{\label{sec:Conclusion}Conclusion}

In this paper we studied incremental load probabilities in the ASIP
model, analyzed their asymptotic behavior and discussed their implications.
Introducing the notion of incremental load, and analyzing it via two
complementary approaches --- a continuum diffusion-limit approach,
and an exact probabilistic-combinatorial approach --- we analytically
derived expressions for the occupation probabilities of the ASIP's
lattice intervals, their corresponding factorial moments, and for
the probability distribution of the ASIP's inter-exit time. Spanning
both exact results and asymptotic behaviors, the analysis presented
herein joins the recently published \cite{ASIP-3} to provide the
most comprehensive description of the ASIP's steady-state statistics
to date. 

Our work is yet another step towards a more profound understanding
of the ASIP's complex dynamical behavior, and is part of a long term
goal --- the elucidation of the ASIP’s steady state distribution in
full detail. As an intermediate step, it is natural to turn to the
study of correlations between the occupations of several disjoint
intervals. The empty interval method was employed in the study of
correlations for ASIPs on a ring \cite{Coalescence Process}, and
may thus also prove useful for open boundary ASIPs. This question
is especially interesting in light of the picture discussed above
of an open ASIP as a `conveyor belt': if a single snapshot of an open
boundary ASIP is similar to the temporal evolution of the ASIP on
a ring, it would be interesting to examine the relation between two-point
correlation functions in the former and two-time correlation functions
in the latter. 

Other interesting questions remain open, many of which are related
to the concept of universality. To this end, it would be very interesting
to examine the robustness, and inevitable collapse, of the results
presented herein with respect to a large range of perturbations. For
example, it would be interesting to further consider the effect of
non-homogeneous hopping rates on cluster formation and delineate the
conditions under which non-homogeneity is asymptotically averaged
out. While some progress in this direction has already been made \cite{ASIP-3},
much still remains to be done. Modifying the ASIP a bit, one may ask
how does a dependence of the hopping rate on cluster size affect the
observed statistics? Another question is what happens when particles
arrive to sites other than the first? Finally, the analysis of a generalized
ASIP in which hopping times are non-exponential would be both interesting
and undoubtedly extremely challenging as it will inevitably require
different methods than the ones applied herein.
\begin{acknowledgments}
We gratefully acknowledge Amir Bar, Or Cohen and David Mukamel for
fruitful discussions. The support of the Israel Science Foundation
(ISF) and of the Minerva Foundation with funding from the Federal
German Ministry for Education and Research is gratefully acknowledged.
Shlomi Reuveni gratefully acknowledges support from the James S. McDonnell
Foundation via its postdoctoral fellowship in studying complex systems.
\end{acknowledgments}
\appendix

\section{\label{sub:Derivation-of-Eq. 409}Derivation of Eq. (\ref{409}) }

In this Appendix we present the derivation of Eq. (\ref{409}). To
do so, we define two auxiliary probability functions{\small{
\begin{equation}
\begin{array}{l}
\hat{P}_{l}^{\text{left}}(t;k,m)\equiv\text{Pr}\Bigl(L(t;k,m)=l\text{ and \ensuremath{X_{k-1}(t)=0}}\Bigr)\\
\\
\hat{P}_{l}^{\text{right}}(t;k,m)\equiv\text{Pr}\Bigl(L(t;k,m)=l\text{ and }X_{k+m}(t)=0\Bigr).\\
\text{ }
\end{array}\label{A1}
\end{equation}
}}These are the probabilities that sites $\{k,\cdots,k+m-1\}$ are
occupied by $l$ particles, and the site immediately to their left/right
is empty. Next, note that the probability that sites $\{k-1,\cdots,k+m-1\}$
support $l$ particles and site $k-1$ is \emph{not} empty is exactly
${P_{l}(t;k-1,m+1)}-\hat{P}_{l}^{\text{left}}(t;k,m)$. Similarly,
the probability that sites $\{k,\cdots,k+m-2\}$ support $l$ particles
and site $k+m-1$ is not empty is exactly $P_{l}(t;k,m-1)-\hat{P}_{l}^{\text{right}}(t;k,m-1)$.
Although the auxiliary probabilities are needed in order to write
down the equation of motion for $P_{l}(t;k,m)$, 
\begin{equation}
\begin{array}{l}
\frac{\partial}{\partial t}P_{l}(t;k,m)=\Bigl(P_{l}(t;k-1,m+1)-\hat{P}_{l}^{\text{left}}(t;k,m)\Bigr)\\
\\
+\Bigl(P_{l}(t;k,m-1)-\hat{P}_{l}^{\text{right}}(t;k,m-1)\Bigr)\\
\\
-\Bigl(P_{l}(t;k,m)-\hat{P}_{l}^{\text{left}}(t;k,m)\Bigr)\\
\\
-\Bigl(P_{l}(t;k,m)-\hat{P}_{l}^{\text{right}}(t;k,m-1)\Bigr)\,,\\
\text{ }
\end{array}\label{A2}
\end{equation}
they cancel out in Eq. (\ref{A2}) and Eq. (\ref{409}) readily follows.

\section{\label{sub:Derivation-of-Eq. 411}Derivation of Eq. (\ref{411})}

The derivation of Eq. (\ref{411}) is similar to the derivation of
Eq. (\ref{409}) albeit replacing terms corresponding to the entry
of particles into the interval from the left (first and third lines
of the right hand side of Eq. (\ref{A2})) with terms corresponding
to the arrival of a particle to the first site. The resulting equation
is

\begin{equation}
\begin{array}{l}
\frac{\partial}{\partial t}P_{l}(t;1,m)=\lambda P_{l-1}(t;1,m)-\lambda P_{l}(t;1,m)\\
\\
+\Bigl(P_{l}(t;1,m-1)-\hat{P}_{l}^{\text{right}}(t;1,m-1)\Bigr)\\
\\
-\Bigl(P_{l}(t;1,m)-\hat{P}_{l}^{\text{right}}(t;1,m-1)\Bigr)\,.\\
\text{ }
\end{array}\label{A3}
\end{equation}
Once again, the auxiliary probabilities cancel out, yielding Eq. (\ref{411}).

\section{Universality of Eqs. (\ref{506}) and (\ref{507}): an explicit example
\label{sec:AppendixUniversality}}

In this Appendix we demonstrate how the asymptotic scaling forms (\ref{506})
and (\ref{507}) emerge for an explicit example of an ASIP with a
generalized arrival process. As explained in Section \ref{sub:ScalingRemarks},
the universality is a result of the central limit theorem for the
distribution of $L(1,k)$, which leads to Eq. (\ref{505}). The scaling
forms are obtained by showing, for the specific example considered
below, that the central limit theorem applies. We also present a formal
argument that heuristically explains why the central limit theorem
is expected to apply for a much larger class of arrival processes. 

Consider an ASIP in which particles may enter the first site not only
one by one, but also in batches of $n=2,3,4,\ldots$ particles. The
arrival of a batch of $n$ particles is assumed to be a Poisson process
with rate $\lambda_{n}$. The occupation of the first site thus increases
according to the rule

\begin{equation}
X_{1},X_{2},\cdots\xrightarrow{\lambda_{n}}X_{1}+n,X_{2},\cdots\,\label{A4}
\end{equation}
{[}this is a generalization of Eq. (\ref{201a}){]}, and otherwise
the ASIP dynamics remains unchanged. The goal of the current calculation
is to find the initial condition $P_{l}(1,m)$ that is generated by
this arrival process, and to analyze the conditions under which the
central limit theorem leads to the approximation (\ref{505}). 

The equation equivalent to (\ref{411}) for this generalized ASIP
is
\begin{equation}
\begin{array}{l}
\frac{\partial}{\partial t}P_{l}(t;1,m)=\Bigl[P_{l}(t;1,m-1)-P_{l}(t;1,m)\Bigr]\\
\\
-\sum_{n=1}^{\infty}\lambda_{n}\Bigl[P_{l}(t;1,m)-P_{l-n}(t;1,m)\Bigr].\\
\text{ }
\end{array}\label{A5}
\end{equation}
 Multiplying by $z^{l}$ and summing over $l$ leads, in the steady
state, to
\begin{equation}
\begin{array}{l}
G(z;1,m-1)-G(z;1,m)=\\
\\
{}[f_{\lambda}(1)-f_{\lambda}(z)]G(z;1,m),\\
\text{ }
\end{array}\label{A6}
\end{equation}
where $f_{\lambda}(z)$ is the generating function for $\lambda_{n}$:
\begin{equation}
f_{\lambda}(z)\equiv\sum_{n=1}^{\infty}\lambda_{n}z^{n}.\label{A7}
\end{equation}
Iterating (\ref{A6}) and using $G(z;1,0)=1$ yields 
\begin{equation}
G(z;1,m)=[1+f_{\lambda}(1)-f_{\lambda}(z)]^{-m}.\label{A8}
\end{equation}
One observes that the distribution of $L(1,m)$ has a product form
and is equal to the distribution of a sum of i.i.d. random variable
whose generating function is 
\begin{equation}
g(z)\equiv[1+f_{\lambda}(1)-f_{\lambda}(z)]^{-1},\label{A9}
\end{equation}
compare with Eq. (\ref{810}). The central limit theorem for this
sum applies when the mean and variance of these i.i.d. variables is
finite, i.e., when $g'(1),g''(1)<\infty$. It is easy to verify that
$g'(1)=f'(1)$ and $g''(1)=2f''(1)+[f'(1)]^{2}$. Thus, as long as
$\sum_{n}n^{2}\lambda_{n}<\infty$, one obtains for $m\gg1$
\begin{equation}
P_{l}(1,m)\sim\delta(l-m\langle\lambda\rangle)=\langle\lambda\rangle^{-1}\delta(m-l/\langle\lambda\rangle),\label{A10}
\end{equation}
where we have defined $\langle\lambda\rangle\equiv\sum_{n}n\lambda_{n}$.

Let us now motivate in a heuristic fashion why the central limit theorem
is expected to hold for a much larger class of arrival processes.
Assume that the arrival process is such that Eq. (\ref{A6}) is replaced
by 
\begin{equation}
G(z;1,m-1)-G(z;1,m)=\mathcal{A}(z)G(z;1,m),\label{A11}
\end{equation}
where $\mathcal{A}(z)$ is a formal notation for the operator associated
with the arrival process. The formal solution of this equation is
$G(z;1,m)=[1+\mathcal{A}(z)]^{-m}$ {[}compare with Eq. (\ref{A8}){]}.
If the operator $[1+\mathcal{A}(z)]^{-1}$ is characterized by a non-vanishing
spectral gap, i.e., there is a finite difference between its largest
and second-largest eigenvalues, then when $m\to\infty$ one has asymptotically
$G(z;1,m)\sim g_{\max}(z)^{m}$, where $g_{\max}(z)$ denotes the
largest eigenvalue of $[1+\mathcal{A}(z)]^{-1}$ for some fixed value
of $z$. If, in addition, $g_{\max}(z)$ is the PGF of a random variable
with finite variance, a central limit theorem holds for $L(1,m)$
and an approximation of the form (\ref{505}) is valid.

\section{\label{sub:Saddle-point-evaluation 604}Saddle point evaluation of
Eq. (\ref{604})}

In this section we show how Eq. (\ref{308}) follows by applying a
saddle point approximation (also known as Laplace's method) to the
sum in Eq. (\ref{604}) in the limit $k\to\infty$. The first step
is to apply Stirling's approximation to the probability density of
the traversal time 
\begin{equation}
\frac{t^{m}e^{-t}}{m!}\simeq\frac{e^{-t+m\log(t/m)+m}}{\sqrt{2\pi m}}\,.\label{A12a}
\end{equation}
Next, we substitute Eqs. (\ref{A12a}) and (\ref{603}) into Eq. (\ref{604})
to obtain
\begin{equation}
P{}_{T_{k}}(t)\simeq\sum_{m=1}^{k-1}\frac{e^{-t+m\log(t/m)+m}}{\sqrt{2\pi m}}\frac{me^{-m^{2}/4(k-m)}}{2(k-m)}\,.\label{A12b}
\end{equation}
Setting $u=m/\sqrt{k}$ we rewrite (\ref{A12b}) as
\begin{equation}
P{}_{T_{k}}(t)\simeq\sum_{u}\frac{e^{-t+u\sqrt{k}\log(t/u\sqrt{k})+u\sqrt{k}}}{\sqrt{2\pi u\sqrt{k}}}\frac{ue^{-u^{2}/4\left(1-u/\sqrt{k}\right)}}{2(\sqrt{k}-u)}\,,\label{A12c}
\end{equation}
where the sum runs over values $u=k^{-1/2},2k^{-1/2},\cdots,k^{1/2}-k^{-1/2}$.
We now observe that 
\begin{equation}
P_{T_{k}}(\sqrt{k}t)\simeq\sum_{u}\frac{k^{1/4}u^{1/2}}{\sqrt{8\pi}(k-\sqrt{k}u)}e^{\sqrt{k}f(u)}\label{A12d}
\end{equation}
with 
\begin{equation}
f(u)\equiv u\log(t/u)+u-t-u^{2}/(4\sqrt{k}-4u)\,.\label{A12e}
\end{equation}

For large $k$, the sum in Eq. (\ref{A12d}) may be approximated by
an integral, which can be evaluated using a saddle point approximation.
We thus search for a saddle point $u^{*}$ for which $f'(u*)=0$ and
find it to be 
\begin{equation}
u^{*}=t-t^{2}/2\sqrt{k}+O(k^{-1})\,,\label{A12f}
\end{equation}
($u^{*}$ is computed to leading in $k$ such that $\underset{k\rightarrow\infty}{\lim}\sqrt{k}f'(u^{*})=0$).
Evaluating the integral approximation of the sum in Eq. (\ref{A12d})
to leading order, we find
\begin{equation}
\begin{array}{l}
P_{T_{k}}(\sqrt{k}t)\simeq\int_{0}^{\sqrt{k}}\frac{k^{3/4}u^{1/2}}{\sqrt{8\pi}(k-\sqrt{k}u)}e^{\sqrt{k}f(u)}du\\
\\
=\frac{te^{-t^{2}/4}}{2\sqrt{k}}+O(k^{-1})\,.\\
\text{ }
\end{array}\label{A12g}
\end{equation}
We now observe that the probability density function of the normalized
inter-exit time $T_{k}/\sqrt{\pi k}$ is related to the probability
density function of $T_{k}$ in the following way 
\begin{equation}
P_{T_{k}/\sqrt{\pi k}}(t)=\sqrt{\pi k}P_{T_{k}}(\sqrt{\pi k}t)\,.\label{A12h}
\end{equation}
Equation (\ref{308}) follows immediately.

\section{\label{sub:Derivation-of-Eq. 704}Derivation of Eq. (\ref{704}) }

Conditioning on the occupancy vector $\mathbf{X}(t)$ and utilizing
the Markovian dynamics of Eq. (\ref{703}) we have

{\footnotesize{
\begin{equation}
\begin{array}{l}
\left\langle z^{L\left(t';1,m\right)}\right\rangle =\left\langle \left\langle z^{L\left(t';1,m\right)}|\mathbf{X}(t)\right\rangle \right\rangle \\
\text{ }\\
=\left\{ \begin{array}{l}
(\lambda\Delta)\left\langle z^{L\left(t;1,m\right)+1}\right\rangle \\
+\\
(\mu_{m}\Delta)\left\langle z^{L\left(t;1,m-1\right)}\right\rangle \\
+\\
\left(1-\left(\lambda+\mu_{m}\right)\Delta\right)\left\langle z^{L\left(t;1,m\right)}\right\rangle \\
+\\
o(\Delta)\,\,.
\end{array}\right.
\end{array}\label{A13}
\end{equation}
}}Equation (\ref{704}) is obtained after rearranging terms in Eq.
(\ref{A13}), dividing by $\Delta$, taking $\Delta\rightarrow0$
and using the PGF notation of Eq. (\ref{702a}).

\section{\label{sub:Derivation-of-Eq. 710}Derivation of Eq. (\ref{710}) }

Conditioning on the occupancy vector $\mathbf{X}(t)$ and utilizing
the Markovian dynamics of Eq. (\ref{709}) we have

{\footnotesize{
\begin{equation}
\begin{array}{l}
\left\langle z^{L\left(t';k,m\right)}\right\rangle =\left\langle \left\langle z^{L\left(t';k,m\right)}|\mathbf{X}(t)\right\rangle \right\rangle \\
\text{ }\\
=\left\{ \begin{array}{l}
(\mu_{k-1}\Delta)\left\langle z^{L\left(t;k-1,m+1\right)}\right\rangle \\
+\\
(\mu_{k+m-1}\Delta)\left\langle z^{L\left(t;k,m-1\right)}\right\rangle \\
+\\
\left(1-(\mu_{k-1}+\mu_{k+m-1})\Delta\right)\left\langle z^{L\left(t;k,l\right)}\right\rangle \\
+\\
o(\Delta)\,\,.
\end{array}\right.
\end{array}\label{A14}
\end{equation}
}}Equation (\ref{710}) is obtained after rearranging terms in Eq.
(\ref{A14}), dividing by $\Delta$, taking $\Delta\rightarrow0$
and using the PGF notation of Eq. (\ref{702a}).

\section{\label{sub:Derivation-of-Eq. 712}Derivation of Eq. (\ref{712}) }

We prove Eq. (\ref{712}) by showing that the probability generating
function $G\left(z;k,m\right)$ it defines satisfies Eq. (\ref{711}).
To this end we apply mathematical induction on the index $k$. We
start by showing that Eq. (\ref{712}) holds for $k=2$ and an arbitrary
value of $m$. Indeed, for $k=2$ Eq. (\ref{712}) reads 
\begin{equation}
\begin{array}{l}
G\left(z;2,m\right)=\Pi\left(2,m\right)\\
\\
\text{ }\\
+\Pi\left(2,m\right)\sum_{j=1}^{m}\frac{\mu_{1}}{\mu_{1}+\mu_{1+j}}\frac{G\left(z;1,j+1\right)}{\Pi\left(2,j\right)}\,.\text{ }
\end{array}\label{A15}
\end{equation}
Substituting Eq. (\ref{A15}) into Eq. (\ref{711}) and utilizing
Eq. (\ref{706}) we have {\small{
\begin{equation}
\begin{array}{l}
\Pi\left(2,m\right)\left(1+\sum_{j=1}^{m}\frac{\mu_{1}}{\mu_{1}+\mu_{1+j}}\frac{\prod_{i=1}^{j+1}\frac{\mu_{i}}{\mu_{i}+\lambda\left(1-z\right)}}{\Pi\left(2,j\right)}\right)\overset{?}{=}\\
\\
+\frac{\mu_{1+m}}{\mu_{1}+\mu_{1+m}}\Pi\left(2,m-1\right)\left(1+\sum_{j=1}^{m-1}\frac{\mu_{1}}{\mu_{1}+\mu_{1+j}}\frac{\prod_{i=1}^{j+1}\frac{\mu_{i}}{\mu_{i}+\lambda\left(1-z\right)}}{\Pi\left(2,j\right)}\right)\\
\\
\text{ }\\
+\frac{\mu_{1}}{\mu_{1}+\mu_{1+m}}\prod_{j=1}^{m+1}\frac{\mu_{j}}{\mu_{j}+\lambda\left(1-z\right)}\,.
\end{array}\label{A16}
\end{equation}
}}Canceling matching terms on both sides of Eq. (\ref{A16}) gives
the trivial identity $0=0$ and proves our claim. 

We finish the proof by showing that if Eq. (\ref{712}) holds for
$k\geq2$ it holds for $k+1$ as well. Indeed, replacing $k$ by $k+1$
in Eqs. (\ref{711}-\ref{712}) we substitute Eq. (\ref{712}) into
Eq. (\ref{711}) and obtain 

{\footnotesize{
\begin{equation}
\begin{array}{l}
\Pi\left(k+1,m\right)\left(1+\sum_{j=1}^{m}\frac{\mu_{k}}{\mu_{k}+\mu_{k+j}}\frac{G\left(z;k,j+1\right)}{\Pi\left(k+1,j\right)}\right)\overset{?}{=}\\
\text{ }\\
\\
+\frac{\mu_{k+m}\Pi\left(k+1,m-1\right)}{\mu_{k}+\mu_{k+m}}\left(1+\sum_{j=1}^{m-1}\frac{\mu_{k}}{\mu_{k}+\mu_{k+j}}\frac{G\left(z;k,j+1\right)}{\Pi\left(k+1,j\right)}\right)\\
\\
+\frac{\mu_{k}\Pi\left(k,m+1\right)}{\mu_{k}+\mu_{k+m}}\left(1+\sum_{j=1}^{m+1}\frac{\mu_{k-1}}{\mu_{k-1}+\mu_{k+j-1}}\frac{G\left(z;k-1,j+1\right)}{\Pi\left(k,j\right)}\right)\,.
\end{array}\label{A18}
\end{equation}
}}Canceling matching terms on both sides gives {\small{
\begin{equation}
\begin{array}{l}
G\left(z;k,m+1\right)=\\
\text{ }\\
\\
\Pi\left(k,m+1\right)\left(1+\sum_{j=1}^{m+1}\frac{\mu_{k-1}}{\mu_{k-1}+\mu_{k+j-1}}\frac{G\left(z;k-1,j+1\right)}{\Pi\left(k,j\right)}\right)
\end{array}\label{A19}
\end{equation}
}}which coincides with Eq. (\ref{712}) for $G\left(z;k,m+1\right)$
and concludes our proof.

\section{Asymptotic analysis of Equation (\ref{808a})\label{sec:Asymptotic-analysis}}

In this Appendix we sketch the asymptotic analysis of the exact expression
for the occupation probabilities (\ref{808a}) and show how the results
of Section \ref{sec:Continuum-limit-of}, and in particular Eqs. (\ref{301}),
(\ref{503c})--(\ref{503d}), and (\ref{507}) can be obtained from
it. We concentrate here solely on the case of $m=1$; the calculation
for other values of $m$ is similar but somewhat more lengthy. 

For the case $m=1,$ the sum in (\ref{808a}) can be rewritten, by
substituting the definition (\ref{804}) and the ``initial condition''
(\ref{501c}), as
\begin{multline}
P_{0}(k,1)=\sum_{i=0}^{k-2}\binom{2i+1}{i}\frac{1}{2i+1}2^{-(2i+1)}+\\
+\sum_{i=0}^{k-2}\binom{k-1+i}{i}\frac{k-1-i}{k-1+i}2^{-(k-1+i)}(1+\lambda)^{-(k-i)}\label{eq:A101}
\end{multline}
for $l=0,$ while for $l\geq1$ it has the form 
\begin{equation}
P_{l}(k,1)=S(2,k),\label{eq:A102}
\end{equation}
where we define 
\begin{align}
S(j_{1},j_{2})\equiv & \Bigl(\frac{\lambda}{1+\lambda}\Bigr)^{l}\sum_{j=j_{1}}^{j_{2}}\frac{j-1}{2k-j-1}\binom{2k-j-1}{k-1}\times\label{eq:A102-1}\\
 & \times\binom{l+j-1}{l}2^{-(2k-j-1)}(1+\lambda)^{-j}.\nonumber 
\end{align}
To obtain these relations we have used the binomial identity $\binom{n-1}{k}-\binom{n}{k}=\frac{n-2k}{n}\binom{n}{k}$. 

We first evaluate the sums in Eq. (\ref{eq:A101}) for large $k$.
The first sum can be calculated exactly, and equals ${1-\Gamma(k-1/2)/\sqrt{\pi}\Gamma(k)}\simeq1-1/\sqrt{\pi k}$.
The main contribution to the second sum is from values of $i$ which
are close to $k$. It can be shown, by expanding the summand for $k\gg k-i$,
that the second sum decays to zero as $k^{-3/2}$ and is therefore
negligible compared to the first. We thus arrive at Eq. (\ref{301}).

We now move on to the asymptotic evaluation of (\ref{eq:A102})--(\ref{eq:A102-1})
for large $k$. The main contribution to the sum, as shown below,
is from values of $j$ which are close to $l.$ Therefore, two cases
are treated separately: (i) $l\ll\sqrt{k},$ and (ii) $l=x\sqrt{k}$
with $x=O(1)$.

Case (i), $l\ll\sqrt{k}$. In this case, the sum is evaluated using
Stirling's approximation
\begin{equation}
2^{-(2k-j-1)}\binom{2k-j-1}{k-1}\simeq\sqrt{\frac{2k-j-1}{2\pi(k-j)(k-1)}}e^{-f_{1}(k)},\label{eq:A103-0}
\end{equation}
with
\begin{align}
f_{1}(k) & =(k-1)\log\frac{k-1}{2k-j-1}+(k-j)\log\frac{k-j}{2k-j-1}\nonumber \\
 & =\frac{j^{2}}{4k}\Bigl[1+O\Bigl(\frac{j}{k}\Bigr)\Bigr].\label{eq:A103}
\end{align}
The term $f_{1}$ in the exponent yields a significant contribution
to the summand only for values of $j$ which are comparable with $\sqrt{k}$,
while for $j\ll\sqrt{k}$ it is negligible. Accordingly, we split
the sum in (\ref{eq:A102}) into two: $P_{l}(k,1)=S(2,N)+S(N+1,k)$,
the first running over $j=2,\ldots,N$, and the second over $j=N+1,\ldots k$.
Here $N=N(k)$ is chosen in such a way that $l\ll N\ll k$. The first
of these sums may be approximated using (\ref{eq:A103-0}) as 
\[
S(2,N)\simeq\Bigl(\frac{\lambda}{1+\lambda}\Bigr)^{l}\sum_{j=2}^{N}\frac{j-1}{\sqrt{4\pi}k^{3/2}}\binom{l+j-1}{l}(1+\lambda)^{-j}.
\]
Since $N\gg1$ and the summand decays exponentially with $j$, replacing
the upper boundary in the last sum by $\infty$ results in a negligible
error. The sum can now be computed exactly, and yields
\[
S(2,N)\simeq\frac{l+1}{\sqrt{4\pi}\lambda^{2}k^{3/2}}.
\]
The contribution of the second sum (from $N+1$ to $k$) is negligible
as long as $l\ll\sqrt{k}.$ To see this, approximate $\binom{l+j-1}{l}\simeq j^{l+1}/l!$
(which is valid for $j>N\gg l$), and then approximate the sum as
an integral:
\begin{multline*}
S(N+1,k)\simeq\\
\simeq\Bigl(\frac{\lambda}{1+\lambda}\Bigr)^{l}\frac{k^{(l-1)/2}}{\sqrt{4\pi}l!}\int_{\frac{N}{\sqrt{k}}}^{\sqrt{k}}y^{l+2}e^{-y^{2}/4-y\sqrt{k}\log(1+\lambda)}dy,
\end{multline*}
where a change of integration variable $y=j/\sqrt{k}$ was made. Once
again, we incur a negligible error by approximating the lower and
upper integration boundaries as $N/\sqrt{k}\simeq0$ and $\sqrt{k}\simeq\infty$.
By evaluating the integral, it can be shown that $S(N+1,k)\ll S(2,N)$,
leading to (\ref{503c})--(\ref{503d}) (remember that here $m=1$). 

Case (ii), $l=x\sqrt{k}$. In this case, since $l\gg1$, one may employ
Stirling's approximation also for the second binomial coefficient
in (\ref{eq:A102-1}). Replacing as before the sum by an integral
with an integration variable $y=j/\sqrt{k}$ leads to 
\[
P_{l}(k,1)\simeq\int_{0}^{\infty}\Bigl(\frac{\lambda}{1+\lambda}\Bigr)^{l}\frac{y^{3/2}}{4\pi k^{3/4}\sqrt{x(x+y)}}e^{-\frac{y^{2}}{4}-\sqrt{k}f_{2}(y)}
\]
with 
\[
f_{2}(y)\equiv x\log\frac{x}{x+y}+y\log\frac{y}{x+y}+y\log(1+\lambda).
\]
For $\sqrt{k}\gg1$, the integral can be evaluated using a saddle
point approximation: $f_{2}$ has a minimum at $y^{*}=x/\lambda$,
where its value is $f_{2}(y^{*})=x\log\lambda/(1+\lambda)$. We therefore
obtain the scaling form
\[
P_{l}(k,1)\simeq\frac{x}{\sqrt{4\pi}\lambda^{2}k}e^{-x^{2}/4\lambda^{2}}
\]
{[}compare with (\ref{507}){]}. Note that this saddle point calculation
carries through to any $1\ll l\ll k$. The results are different,
however, at the scale of $l=O(k)$, as the main contribution to the
sum (the saddle point) comes from values $j=O(k)$, leading to non-negligible
corrections to the calculation due to terms neglected above such as
the higher order terms in (\ref{eq:A103}).

\section{\label{sub:Derivation-of-Eq. 811}Derivation of Eq. (\ref{816}) }

In this Appendix we provide an alternative derivation of Eq. (\ref{816}).
The derivation of this Appendix is algebraic in nature and serves
to show that the desired result may also be obtained without reference
to the probabilistic argumentation presented in the main text. The
proof is divided into three parts. In Part I we show that for $k>1$,
$G\left(z;k,m\right)$ can be written as 
\begin{equation}
\begin{array}{l}
G\left(z;k,m\right)=\overset{k}{\underset{j=2}{\sum}}\left(\frac{1}{2}\right)^{2k+m-2j}C_{1}(k+m-j-1,k-j)\\
\text{ }\\
+\left(\frac{1}{2}\right)^{2k+m-2}\overset{\infty}{\underset{l=0}{\sum}}A(k,m,l)z^{l}\text{ }
\end{array}\label{A20}
\end{equation}
where 
\begin{equation}
\begin{array}{l}
A(k,m,l)=\left(\frac{\lambda}{1+\lambda}\right)^{l}\sum_{j_{1}=1}^{m}\sum_{j_{2}=1}^{j_{1}+1}\sum_{j_{3}=1}^{j_{2}+1}\cdot\cdot\cdot\\
\text{ }\\
\cdot\cdot\cdot\sum_{j_{k-2}=1}^{j_{k-3}+1}\sum_{j_{k-1}=1}^{j_{k-2}+1}\binom{j_{k-1}+l}{j_{k-1}}\left(\frac{2}{1+\lambda}\right)^{j_{k-1}+1}\text{ .}
\end{array}\label{A21}
\end{equation}
In Part two we show that{\small{
\begin{equation}
\begin{array}{l}
A(k,m,l)=\\
\\
\text{ }\\
\sum_{j=1}^{m+k-2}2^{j+1}C_{m}(k-2,m+k-2-j)P_{l}(1,j+1)\,.
\end{array}\label{A22}
\end{equation}
}}In Part three we combine Eqs. (\ref{A20}) and (\ref{A22}) to conclude
the proof.

\subsubsection{Part I}

We prove Eq. (\ref{A20}) by induction on $k$. We start by showing
that Eq. (\ref{A20}) holds for $k=2$ and an arbitrary value of $m$.
Setting $\mu_{i}=\mu$ ($i=1,2,3,\cdots$) in Eq. (\ref{712}) we
have 
\begin{equation}
\begin{array}{l}
G\left(z;k,m\right)=\left(\frac{1}{2}\right)^{m}\\
\text{ }\\
+\sum_{j=1}^{m}\left(\frac{1}{2}\right)^{m+1-j}G\left(z;k-1,j+1\right)\text{ }
\end{array}\label{A23}
\end{equation}
($k>1$). Setting $k=2$ in Eq. (\ref{A23}) and utilizing Eq. (\ref{810})
we have

\begin{equation}
\begin{array}{l}
G\left(z;2,m\right)=\left(\frac{1}{2}\right)^{m}\\
\text{ }\\
\\
+\sum_{j=1}^{m}\left(\frac{1}{2}\right)^{m+1-j}\left(\frac{1}{1+\lambda\left(1-z\right)}\right)^{j+1}\text{ .}
\end{array}\label{A24}
\end{equation}
Recalling the Taylor expansion 
\begin{equation}
\frac{1}{1-x}=\overset{\infty}{\underset{i=0}{\sum}}x^{i}\label{A25}
\end{equation}
$|x|<1$, we expand the parenthesis in the second term of Eq. (\ref{A24})
to obtain
\begin{equation}
\begin{array}{l}
G\left(z;2,m\right)=\left(\frac{1}{2}\right)^{m}\\
\text{ }\\
\\
+\sum_{j=1}^{m}\left(\frac{1}{2}\right)^{m+1-j}\left(\frac{1}{1+\lambda}\right)^{j+1}\left(\overset{\infty}{\underset{i=0}{\sum}}\left(\frac{\lambda z}{1+\lambda}\right)^{i}\right)^{j+1}\text{ .}
\end{array}\label{A26}
\end{equation}
Noting that 
\begin{equation}
\left(\overset{\infty}{\underset{i=0}{\sum}}\left(\frac{\lambda z}{1+\lambda}\right)^{i}\right)^{j+1}=\overset{\infty}{\underset{l=0}{\sum}}\left(\begin{array}{c}
j+l\\
j
\end{array}\right)\left(\frac{\lambda z}{1+\lambda}\right)^{l}\label{A27}
\end{equation}
and 
\begin{equation}
A(2,m,l)=\left(\frac{\lambda}{1+\lambda}\right)^{l}\sum_{j=1}^{m}\left(\frac{2}{1+\lambda}\right)^{j+1}\left(\begin{array}{c}
j+l\\
j
\end{array}\right)\label{A28}
\end{equation}
we substitute Eq. (\ref{A27}) into Eq. (\ref{A26}) to obtain 
\begin{equation}
G\left(z;2,m\right)=\left(\frac{1}{2}\right)^{m}+\left(\frac{1}{2}\right)^{m+2}\overset{\infty}{\underset{l=0}{\sum}}A(2,m,l)z^{l}\text{ .}\label{A29}
\end{equation}
Noting that $C_{1}\left(m-1,0\right)=1$ ($m=1,2,3,\cdots)$, we see
that Eq. (\ref{A29}) identifies with (\ref{A20}) for $k=2$.

We finish the first part of the proof by showing that if Eq. (\ref{A20})
holds for $k\geq2$ it holds for $k+1$ as well. Indeed, replacing
$k$ by $k+1$ in Eq. (\ref{A23}) we substitute Eq. (\ref{A20})
into Eq. (\ref{A23}) and obtain
\begin{equation}
\begin{array}{l}
G\left(z;k+1,m\right)=\\
\\
+\left(\frac{1}{2}\right)^{m}\left[1+\overset{m}{\underset{i=1}{\sum}}\overset{k}{\underset{j=2}{\sum}}\left(\frac{1}{2}\right)^{2k+2-2j}C_{1}(k-j+i,k-j)\right]\\
\\
\text{ }\\
+\left(\frac{1}{2}\right)^{m+2k}\overset{m}{\underset{i=1}{\sum}}\overset{\infty}{\underset{l=0}{\sum}}A(k,i+1,l)z^{l}\text{ .}
\end{array}\label{A30a}
\end{equation}
Performing an index shift $j\rightarrow k+1-j$, Eq. (\ref{A30a})
can be rewritten as{\small{ }}
\begin{equation}
\begin{array}{l}
G\left(z;k+1,m\right)=\\
\\
+\left(\frac{1}{2}\right)^{m}\left[1+\overset{m}{\underset{i=1}{\sum}}\overset{k-1}{\underset{j=1}{\sum}}\left(\frac{1}{2}\right)^{2j}C_{1}(i+j-1,j-1)\right]\\
\\
\text{ }\\
+\left(\frac{1}{2}\right)^{m+2k}\overset{m}{\underset{i=1}{\sum}}\overset{\infty}{\underset{l=0}{\sum}}A(k,i+1,l)z^{l}\text{ .}
\end{array}\label{A30b}
\end{equation}
We now note that Eqs. (\ref{803}) and (\ref{A21}) imply respectively
that 
\begin{equation}
C_{1}(j+m-1,j)=\sum_{i=1}^{m}C_{1}(i+j-1,j-1)\label{A31}
\end{equation}
and
\begin{equation}
A(k+1,m,l)=\sum_{i=1}^{m}A(k,i+1,l)\,.\label{A32}
\end{equation}
Substituting Eqs. (\ref{A31}) and (\ref{A32}) into Eq. (\ref{A30b})
we obtain 
\begin{equation}
\begin{array}{l}
G\left(z;k+1,m\right)=\\
\\
+\left(\frac{1}{2}\right)^{m}\left[1+\overset{k-1}{\underset{j=1}{\sum}}\left(\frac{1}{2}\right)^{2j}C_{1}(j+m-1,j)\right]\\
\\
\text{ }\\
+\left(\frac{1}{2}\right)^{m+2k}\overset{\infty}{\underset{l=0}{\sum}}A(k+1,m,l)z^{l}\text{ .}
\end{array}\label{A33}
\end{equation}
Applying the index shift $j\rightarrow k-j+1$ and noting again that
$C_{1}(m-1,0)=1$ ($m=1,2,3,\cdots)$ we conclude that
\begin{equation}
\begin{array}{l}
G\left(z;k+1,m\right)=\\
\\
+\overset{k+1}{\underset{j=2}{\sum}}\left(\frac{1}{2}\right)^{2k+m+2-2j}C_{1}(k+m-j,k+1-j)\\
\\
\text{ }\\
+\left(\frac{1}{2}\right)^{m+2k}\overset{\infty}{\underset{l=0}{\sum}}A(k+1,m,l)z^{l}\text{ ,}
\end{array}\label{A34}
\end{equation}
a form which coincides with Eq. (\ref{A20}) for $G\left(z;k+1,m\right)$.

\subsubsection{Part II}

We will now prove Eq. (\ref{A22}). Examining Eq. (\ref{A21}) it
is easy to see that it can be rewritten in the following form
\begin{equation}
\begin{array}{l}
A(k,m,l)=\sum_{j=1}^{m+k-2}2^{j+1}\#_{k,j}^{m}P_{l}(1,j+1)\,,\\
\text{ }
\end{array}\label{A36}
\end{equation}
where we have used Eq. (\ref{501c}) and defined{\small{ }}
\begin{equation}
\begin{array}{l}
\#_{k,j}^{m}=\sum_{j_{1}=1}^{m}\sum_{j_{2}=1}^{j_{1}+1}\sum_{j_{3}=1}^{j_{2}+1}\cdot\cdot\cdot\\
\text{ }\\
\cdot\cdot\cdot\sum_{j_{k-2}=1}^{j_{k-3}+1}\sum_{j_{k-1}=1}^{j_{k-2}+1}\delta(j_{k-1},j)\text{ }
\end{array}\label{A37}
\end{equation}
to be the exact number of times that the running index $j_{k-1}$
in Eq. (\ref{A21}) is equal to $j$ ($j=1,\cdots,m+k-2$).

What can be said about the numbers $\#_{k,j}^{m}$? First, it is fairly
straightforward to see that when $k=2$ we have 
\begin{equation}
\begin{array}{l}
\#_{2,j}^{m}=1\end{array}\label{A38}
\end{equation}
($m=1,2,\cdots$;$j=1,\cdots,m$). In addition when $j=m+k-2$ we
have 
\begin{equation}
\begin{array}{l}
\#_{k,m+k-2}^{m}=1\end{array}\label{A39}
\end{equation}
($m=1,2,\cdots$;$k=2,3,\cdots$). Now, for $k>2$ and $1\leq j<m+k-2$
we note that the following recursion relation holds 
\begin{equation}
\begin{array}{l}
\#_{k,j}^{m}=\#_{k-1,j-1}^{m}+\#_{k,j+1}^{m}\,.\end{array}\label{A40}
\end{equation}
Indeed, substituting Eq. (\ref{A37}) into Eq. (\ref{A40}) we have
{\small{
\begin{equation}
\begin{array}{l}
\#_{k,j}^{m}\overset{?}{=}\sum_{j_{1}=1}^{m}\sum_{j_{2}=1}^{j_{1}+1}\sum_{j_{3}=1}^{j_{2}+1}\cdot\cdot\cdot\sum_{j_{k-2}=1}^{j_{k-3}+1}\delta(j_{k-2},j-1)\\
\text{ }\\
\text{ }\\
+\sum_{j_{1}=1}^{m}\sum_{j_{2}=1}^{j_{1}+1}\sum_{j_{3}=1}^{j_{2}+1}\cdot\cdot\cdot\sum_{j_{k-2}=1}^{j_{k-3}+1}\sum_{j_{k-1}=1}^{j_{k-2}+1}\delta(j_{k-1},j+1)
\end{array}\label{A41}
\end{equation}
}}which immediately gives
\begin{equation}
\begin{array}{l}
\#_{k,j}^{m}\overset{?}{=}\sum_{j_{1}=1}^{m}\sum_{j_{2}=1}^{j_{1}+1}\sum_{j_{3}=1}^{j_{2}+1}\cdot\cdot\cdot\\
\text{ }\\
\cdot\cdot\cdot\sum_{j_{k-2}=1}^{j_{k-3}+1}\left[\delta(j_{k-2},j-1)+\sum_{j_{k-1}=1}^{j_{k-2}+1}\delta(j_{k-1},j+1)\right]\text{ }
\end{array}\label{A42}
\end{equation}
However, it easy to check that 
\begin{equation}
\sum_{j_{k-1}=1}^{j_{k-2}+1}\delta(j_{k-1},j+1)=\left(\sum_{j_{k-1}=1}^{j_{k-2}+1}\delta(j_{k-1},j)\right)-\delta(j_{k-2},j-1)\label{A43}
\end{equation}
substituting Eq. (\ref{A43}) into Eq. (\ref{A42}) we recover Eq.
(\ref{A37}) and assert the validity of Eq. (\ref{A40}). 

We now note that
\begin{equation}
\#_{k,j}^{m}=C_{m}(k-2,m+k-2-j)\,.\label{A44}
\end{equation}
Indeed, for $k=2$ 
\begin{equation}
\begin{array}{l}
\#_{2,j}^{m}=C_{m}(0,m-j)=1\end{array}\label{A45}
\end{equation}
($m=1,2,\cdots$;$j=1,\cdots,m$). In addition, for $j=m+k-2$ we
have 
\begin{equation}
\begin{array}{l}
\#_{k,m+k-2}^{m}=C_{m}(k-2,0)=1\end{array}\label{A46}
\end{equation}
($m=1,2,\cdots$;$k=2,3,\cdots$). Finally we note that, for $k>2$
and $1\leq j<m+k-2$, Eqs. (\ref{A40}) and (\ref{A44}) imply that
\begin{equation}
\begin{array}{l}
C_{m}(k-2,m+k-2-j)=C_{m}(k-3,m+k-2-j)\\
\text{ }\\
+C_{m}(k-2,m+k-3-j)\text{ }
\end{array}\label{A47}
\end{equation}
which together with the boundary conditions specified in Eqs. (\ref{A45}--\ref{A46})
give back the iterative construction of the Catalan trapezoid of order
$m$. Substituting Eq. (\ref{A44}) into Eq. (\ref{A36}) we recover
prove Eq. (\ref{A22}) and conclude the second part our proof.

\subsubsection{Part III}

In this part we complete the derivation of Eq. (\ref{816}). Substituting
Eq. (\ref{A22}) into Eq. (\ref{A20}) we have{\scriptsize{
\begin{equation}
\begin{array}{l}
G\left(z;k,m\right)=\overset{k}{\underset{j=2}{\sum}}\left(\frac{1}{2}\right)^{2k+m-2j}C_{1}(k+m-j-1,k-j)\\
\text{ }\\
+\overset{\infty}{\underset{l=0}{\sum}}\left[\overset{m+k-2}{\underset{j=1}{\sum}}P_{l}(1,j+1)\cdot\left(\frac{1}{2}\right)^{2k+m-3-j}C_{m}(k-2,m+k-2-j)\right]z^{l}\text{ }
\end{array}\label{A48}
\end{equation}
}}where we have utilized the fact that $P_{0}(j,0)=1$. Shifting the
index of summation in the inner sum of the second line of Eq. (\ref{A48})
we obtain (\ref{816}).


\begin{thebibliography}{References}
\bibitem{ASIP-1}S. Reuveni, I. Eliazar and U. Yechiali, Asymmetric
Inclusion Process. \emph{Phys. Rev. E.} \textbf{84}, 041101, (2011). 

\bibitem{Grosskinsky}Note that the name ASIP was also used for a
different model: S. Grosskinsky, F. Redig and K. Vafayi, Condensation
in the inclusion process and related models. Journal of Statistical
Physics, Journal of Statistical Physics, \textbf{142}, 5, 952-974,
(2011). 

\bibitem{Derrida1}B. Derrida, E. Domany, and D. Mukamel. An exact
solution of the one dimensional asymmetric exclusion model with open
boundaries, \textit{Journal of Statistical Physics} \textbf{69}, 667
(1992).

\bibitem{Golinelli}O. Golinelli and K. Mallick, The asymmetric simple
exclusion process: an integrable model for non-equilibrium statistical
mechanics, \emph{J. Phys. A}: Math. Gen. \textbf{39}, 12679, (2006). 

\bibitem{Derrida2}B. Derrida, Non-equilibrium steady states: fluctuations
and large deviations of the density and of the current,\textit{ J.
Stat. Mech.} P07023 (2007).

\bibitem{Jackson1} J. R. Jackson, \textit{Operations Research} \textbf{5},
518, (1957).

\bibitem{Jackson2} J. R. Jackson, \textit{Management Science} \textbf{10},
131, (1963).

\bibitem{Fundamentals of Queueing Networks}H. Chen and D. D. Yao,\textit{
Fundamentals of Queueing Networks, }Springer, (2001).

\bibitem{Neuts}M. F. Neuts, The Busy Period of a Queue with Batch
Service. \emph{Operations Research} \textbf{13} 815-819 (1965).

\bibitem{Kaspi}H. Kaspi, O. Kella and D. Perry, Dam processes with
state dependent batch sizes and intermittent production processes
with state dependent rates. \emph{Queueing Systems: Theory and Applications}
\textbf{24} 37- 57 (1997).

\bibitem{van der Wal1}J. van der Wal and U. Yechiali, Dynamic Visit-Order
Rules for Batch-Service Polling. \emph{Probability in the Engineering
and Informational Sciences}, \textbf{17}, No. 3, 351, (2003)

\bibitem{van der Wal2}O. Boxma, J. van der Wal and U. Yechiali, Polling
with Gated Batch Service. \emph{Proceedings of the Sixth International
Conference on \textquotedbl{}Analysis of Manufacturing Systems\textquotedbl{}},
Lunteren, The Netherlands, 155, (2007).

\bibitem{van der Wal3}O. Boxma, J. van der Wal and U. Yechiali, Polling
with Batch Service. \emph{Stochastic Models}, \textbf{24}, No.4, 604,
(2008).

\bibitem{ASIP-2}S. Reuveni, I. Eliazar and U. Yechiali, The Asymmetric
Inclusion Process: A Showcase of Complexity. \emph{Phys. Rev. Lett.
}\textbf{109}, 020603, (2012).

\bibitem{ASIP-3}S. Reuveni, I. Eliazar and U. Yechiali, Limit Laws
for the Asymmetric Inclusion Process. \emph{Phys. Rev. E.} \textbf{86},
061133, (2012). 

\bibitem{Kinetic View}P. L. Krapivsky, S. Redner, and E. Ben-Naim.
A kinetic view of statistical physics. Cambridge University Press,
Cambridge, UK, 2010.

\bibitem{Coalescence Process}D. Ben-Avraham. The coalescence process,
$A+A\rightarrow A$, and the method of interparticle distribution
functions. In V. Privman, editor, Nonequilibrium Statistical Mechanics
in One Dimension, pages 29–50. Cambridge University Press, Cambridge,
UK, 2005.

\bibitem{Catalan Trapezoid}S. Reuveni, Catalan's Trapezoids. Under
Review.

\bibitem{Catalan}K. Thomas, Catalan Numbers with Applications, Oxford
University Press, ISBN 0-19-533454-X (2008).

\bibitem{Catalan's_triangle1}H.G. Forder, Some Problems in Combinatorics.
\emph{Math. Gaz.} \textbf{45}, 199, (1961).

\bibitem{Catalan's_triangle2}L. W. Shapiro, A Catalan Triangle. \emph{Disc.
Math.} \textbf{14}, 83, (1976).

\bibitem{Catalan's_triangle3}D. F. Bailey, Counting Arrangements
of 1's and -1's. \emph{Math. Mag.} \textbf{69}, 128, (1996).

\bibitem{Jain}K. Jain and M. Barma, Phases of a conserved model of
aggregation with fragmentation at fixed sites. \emph{Phys. Rev. E.}
\textbf{64}, 016107, (2001).

\bibitem{Smoluchowski}M., Smoluchowski, Versuch einer mathematischen
Theorie der Koagulationskinetik kolloider Lösungen. \emph{Z. phys.
Chem.} \textbf{92}, (1917).

\bibitem{Sokolov}I. M. Sokolov, S. B. Yuste, J. J. Ruiz-Lorenzo,
and K. Lindenberg. Mean fi{}eld model of coagulation and annihilation
reactions in a medium of quenched traps: Subdiffusion. \emph{Phys.
Rev. E} \textbf{80} , 051114, (2009).

\bibitem{Lindenberg}S. B. Yuste, J. J. Ruiz-Lorenzo, and K. Lindenberg.
Coagulation reactions in low dimensions: Revisiting subdiffusive A+A
reactions in one dimension. \emph{Phys. Rev. E} \textbf{80} , 051114,
(2009).

\bibitem{Derrida3}B. Derrida, V. Hakim, V. Pasquier, Exact First-Passage
Exponents of 1D Domain Growth: Relation to a Reaction-Diffusion Model.
\emph{Phys. Rev. Lett.} \textbf{75}, 751, (1995). 

\bibitem{Blythe}R. A. Blythe and M. R., Evans, Nonequilibrium steady
states of matrix-product form: a solver’s guide. \textit{J. Phys.
A: Math. Theor.} \textbf{40,} R333-R441, (2007). 

\bibitem{MacDonald}C. T. MacDonald, J. H. Gibbs and A. C. Pipkin,
Kinetics of biopolymerization on nucleic acid templates, \textit{Biopolymers}
\textbf{6,} 1, (1968).

\bibitem{Spitzer}F. Spitzer, Interaction of Markov Processes. \emph{Advances
in Mathematics} \textbf{5}, 246, (1970).

\bibitem{Heckmann}K. Heckmann, Single file diffusion Passive Permeability
of Cell Membranes. \textit{Biomembranes} \textbf{3} 127 ed. Kreuzer
F. and Slegers J. F. G., New York: Plenum, (1972).

\bibitem{Levitt}D. G. Levitt, Dynamics of a single-file pore: Non-Fickian
behavior, \emph{Phys. Rev. A} \textbf{8} 3050 (1973).

\bibitem{Richards}P. M. Richards, Theory of one-dimensional hopping
conductivity and diffusion, \emph{Phys. Rev. B} \textbf{16} 1393 (1977).

\bibitem{Widom}B. Widom B, J. L. Viovy and A. D. Defontaines, Repton
model of gel electrophoresis and diffusion, \emph{J. Physique I} \textbf{1}
1759 (1991).

\bibitem{Schreckenberg}M. Schreckenberg and D. E. Wolf (ed) Traffic
and Granular Flow ’97 (New York: Springer) (1998).

\bibitem{Shaw} L. B. Shaw, R.K. Zia and K.H. Lee, Totally asymmetric
exclusion process with extended objects: a model for protein synthesis.
\emph{Phys Rev E} \textbf{68} 021910 (2003).

\bibitem{Reuveni}S. Reuveni, I. Meilijson, M. Kupiec, E. Ruppin and
T. Tuller, Genome-Scale Analysis of Translation Elongation with a
Ribosome Flow Mode. PLoS Computational Biology 7(9), e1002127, (2011).

\bibitem{Halpin}T. Halpin-Healy and Y. C. Zhang, Kinetic roughening
phenomena, stochastic growth, directed polymers and all that, \emph{Phys.
Rep.} \textbf{254} 215 (1995).

\bibitem{Krug}J. Krug, Origins of scale invariance in growth processes,
\emph{Adv. Phys.} \textbf{46} 139 (1997).

\bibitem{Bundschuh}R. Bundschuh, Asymmetric exclusion process and
extremal statistics of random sequences, \emph{Phys. Rev. E} \textbf{65}
031911 (2002).

\bibitem{Klumpp}S. Klumpp S and R. Lipowsky, Traffic of molecular
motors through tube-like compartments, \emph{J. Stat. Phys.} \textbf{113}
233 (2003).

\bibitem{Oshanin1}S. F. Burlatsky, G. S. Oshanin, A. V. Mogutov and
M. Moreau, Directed walk in a one-dimensional lattice gas. \emph{Physics
Letters A }\textbf{166} 230-234 (1992).

\bibitem{Oshanin2}S. F. Burlatsky, G. Oshanin, M. Moreau and W. P.
Reinhardt, Motion of a driven tracer particle in a one-dimensional
symmetric lattice gas. \emph{Phys Rev E} \textbf{54} 3165-3172 (1996).

\bibitem{Oshanin3}O. Bénichou, A. M. Cazabat, J. De Coninck, M. Moreau
and G. Oshanin, Stokes Formula and Density Perturbances for Driven
Tracer Diffusion in an Adsorbed Monolayer. \emph{Phys. Rev. Lett.}
\textbf{84} 511-514 (2000).

\bibitem{Monasterio}C. M. Monasterio and Gleb Oshanin, Bias- and
bath-mediated pairing of particles driven through a quiescent medium.
\emph{Soft Matter} \textbf{7} 993-1000 (2011).

\bibitem{Fundamentals of Queueing Theory}D. Gross, J. F. Shortle,
J.M. Thompson amd C.M. Harris,\emph{ Fundamentals of Queueing Theory,
}Wiley-Interscience, (2008).

\bibitem{Telecommunications1}J. N. Daigle, Queueing Theory with Applications
to Packet Telecommunication, Springer, (2004).

\bibitem{Telecommunications2}G. Giambene, Queuing Theory and Telecommunications:
Networks and Applications, Springer, (2004).

\bibitem{Telecommunications3}A. S. Alfa, Queueing Theory for Telecommunications:
Discrete Time Modeling of a Single Node System, Springer, (2010). 

\bibitem{Traffic Engineering}G. Ausgabe, Traffic Theory, Kluwer,
(2002). 

\bibitem{performance1}G. Bolch, S. Greiner, H. de Meer and K. S.
Trivedi, Queueing Networks and Markov Chains: Modeling and Performance
Evaluation with Computer Science Applications, Wiley-Interscience,
2nd edition, (2006).

\bibitem{performance2}W. J. Stewart, Probability, Markov Chains,
Queues, and Simulation: The Mathematical Basis of Performance Modeling,
Princeton University Press, (2009).

\bibitem{performance3}R. Nelson, Probability, Stochastic Processes,
and Queueing Theory: The Mathematics of Computer Performance Modeling,
Springer, (2010).

\bibitem{human behavior1}A. L. Barabási, The origin of bursts and
heavy tails in human dynamics. \emph{Nature}, \textbf{435}, 207, (2005)

\bibitem{human behavior2}A. Vázquez, Exact results for the Barabási
model of human dynamics. \emph{Phys. Rev. Lett.}, \textbf{95}, 248701,
(2005).

\bibitem{human behavior3}J. Walraevens, T. Demoor, T. Maertens and
H. Bruneel, Stochastic queueing-theory approach to human dynamics.
\emph{Phys. Rev. E,} \textbf{85}, 021139, (2012).

\bibitem{human behavior4}Vidar Frette and P. C. Hemmer, Time needed
to board an airplane: A power law and the structure behind it. \emph{Phys.
Rev. E,} \textbf{85}, 011130, (2012).

\bibitem{Arazi1}A. Arazi, E. Ben-Jacob and U. Yechiali, Bridging
Genetic Networks and Queueing Theory. \emph{Physica A,} \textbf{332},
585, (2004).

\bibitem{Arazi2}A. Arazi and U. Yechiali, Modeling Genetic Regulatory
Systems via a G-Network. Technical Report (2004). 

\bibitem{gene expression1}V. Elgart, T. Jia and R. V. Kulkarni, Applications
of Little’s Law to stochastic models of gene expression. \emph{Phys.
Rev. E}, \textbf{82}, 021901, (2010).

\bibitem{gene expression2}T. Jia and R. V. Kulkarni, Intrinsic Noise
in Stochastic Models of Gene Expression with Molecular Memory and
Bursting. \emph{Phys. Rev. Lett.,} \textbf{106}, 058102, (2011).

\bibitem{microtubule}P. Ashwin, C. Lin and G. Steinberg, Queueing
induced by bidirectional motor motion near the end of a microtubule.
\emph{Phys. Rev. E,} \textbf{82}, 051907, (2010).

\bibitem{Sandpiles}B. Tadić and V. Priezzhev, Scaling of avalanche
queues in directed dissipative sandpiles.\emph{ Phys. Rev. E,} \textbf{62},
3266, (2000).

\bibitem{Chernyak1}Y. Chernyak, M. Chertkov, D. A. Goldberg and K.
Turitsyn, Non-Equilibrium Statistical Physics of Currents in Queuing
Networks. \emph{J. Stat Phys,} \textbf{140}, 819, (2010).

\bibitem{Chernyak2}V. Y. Chernyak, M. Chertkov and N. A. Sinitsyn,
The geometric universality of currents. \emph{J. Stat. Mech.}, P09006,
(2011). 

\bibitem{Kafri}N. Merhav and Y. Kafri, Bose-Einstein condensation
in large deviations with applications to information systems. \emph{J.
Stat. Mech.,} P02011, (2010).

\bibitem{Arita1}C. Arita, Queueing process with excluded-volume effect.
\emph{Phys. Rev. E,} \textbf{80}, 051119, (2009).

\bibitem{Arita2}C. Arita, Dynamical analysis of the exclusive queueing
process, \emph{Phys. Rev. E}, \textbf{83}, 051128 (2011).

\bibitem{Arita3}C. Arita and A. Schadschneider, Exact dynamical state
of the exclusive queueing process with deterministic hopping. \emph{Phys.
Rev. E}, \textbf{84}, 051127, (2011).

\bibitem{Barenblatt}G. I. Barenblatt, \textit{Scaling, self-similarity,
and intermediate asymptotics}, Cambridge, Cambridge University Press,
(1996).

\bibitem{Abramowitz}M. Abramowitz, I.A. Stegun, Handbook of Mathematical
Functions with Formulas, Graphs, and Mathematical Tables, New York,
Dover Publications, (1972).

\bibitem{Images}S. Redner, \emph{A Guide to First-Passage Processes,}
Cambridge University Press, (2001).

\bibitem{Wolff}R. W. Wolff, Stochastic Modeling and the Theory of
Queues, Prentice-Hall, (1989).\end{thebibliography}
\end{document}